\documentclass[journal]{IEEEtran}
\usepackage[normalem]{ulem}
\usepackage[noadjust]{cite}
\usepackage{bbm}
\usepackage{array}
\usepackage{dcolumn}
\usepackage{epsfig}
\usepackage[intlimits]{amsmath}
\usepackage{amsmath, amsfonts, yhmath, bm}
\usepackage{amssymb}
\usepackage{psfrag}
\usepackage{color,soul}
\usepackage[dvipsnames]{xcolor}
\usepackage[normalem]{ulem}
\usepackage{enumerate}
\usepackage{stackengine}
\usepackage[noadjust]{cite}
\usepackage{graphicx}
\usepackage{caption}
\usepackage{subcaption}
\usepackage[font=footnotesize]{subcaption}
\usepackage{multirow}
\usepackage{cite}
\usepackage[font=footnotesize]{caption}
\usepackage{etoolbox}
\usepackage{tcolorbox}
\usepackage{float}
\usepackage{dsfont}
\usepackage[ruled,vlined]{algorithm2e}
\usepackage{algpseudocode}
\usepackage{tikz}
\usepackage{mathtools}
\usetikzlibrary{arrows}
\definecolor{lgreen} {RGB}{180,210,100}
\definecolor{ngreen} {RGB}{98,158,31}
\definecolor{dgreen} {RGB}{78,138,21}
\definecolor{MLOWLSgreen} {RGB}{0,140,130}
\definecolor{SDPpurple} {RGB}{191,0,191}
\definecolor{lred}   {RGB}{220,0,0}
\definecolor{nred}   {RGB}{224,0,0}
\definecolor{bred}   {RGB}{200,20,20}
\definecolor{nblue}  {RGB}{28,130,185}
\definecolor{jblue}  {RGB}{20,50,100}

\definecolor{bred}   {RGB}{200,20,20}
\definecolor{crimson} {RGB}{220,20,62}
\newcommand*\circled[1]{\tikz[baseline=(char.base)]{
		\node[circle,draw,color=crimson, opacity=0.9,inner sep=1pt] (char) {\footnotesize #1};}}
\newcommand*\circledsmall[1]{\tikz[baseline=(char.base)]{
		\node[circle,draw,color=nblue, opacity=0.9,inner sep=1pt] (char) {\scriptsize #1};}}
\newcommand*\circledjblue[1]{\tikz[baseline=(char.base)]{
		\node[circle,draw,color=nblue, opacity=0.9,inner sep=1pt] (char) {\footnotesize #1};}}

\newcommand {\myvec}[1] {{\mbox{\boldmath $#1$}}}

\newcommand*{\myfontb}{\fontfamily{lmr}\selectfont}

\algnewcommand{\algorithmicforeach}{\textbf{for each}}
\algdef{SE}[FOR]{ForEach}{EndForEach}[1]
{\algorithmicforeach\ #1\ \algorithmicdo}
{\algorithmicend\ \algorithmicforeach}

\DeclareMathAlphabet      {\mathbfit}{OML}{cmm}{b}{it}
\DeclareMathAlphabet	  {\mathbfcal}{OMS}{cmsy}{b}{n}

\newcommand {\uepsilon} {\myvec{\varepsilon}}

\newcommand {\ux} {\myvec{x}}

\newcommand {\ubv} {\mybar{\myvec{v}}}

\newcommand {\uv} {\myvec{v}}

\newcommand {\uh} {\myvec{h}}

\newcommand {\uone} {\myvec{1}}

\newcommand {\Rset} {\mathbb{R}}

\newcommand {\Eset} {\mathbb{E}}
\newcommand {\Nset} {\mathbb{N}}

\newcommand {\Tr} {\text{\normalfont Tr}}
\newcommand {\tps} {\rm{T}}

\newcommand {\nooverloadnp} { \mathcal{E}_{\tiny \mybar{{\rm{OL}}}_n}^{(p)} }
\newcommand {\nooverloadnn} { \mathcal{E}_{\tiny \mybar{{\rm{OL}}}_{n+1}}^{(n)} }

\makeatletter
\newsavebox\myboxA
\newsavebox\myboxB
\newlength\mylenA

\newcommand*\mybar[2][0.75]{%
	\sbox{\myboxA}{$\m@th#2$}%
	\setbox\myboxB\null
	\ht\myboxB=\ht\myboxA%
	\dp\myboxB=\dp\myboxA%
	\wd\myboxB=#1\wd\myboxA
	\sbox\myboxB{$\m@th\overline{\copy\myboxB}$}
	\setlength\mylenA{\the\wd\myboxA}
	\addtolength\mylenA{-\the\wd\myboxB}%
	\ifdim\wd\myboxB<\wd\myboxA%
	\rlap{\hskip 0.5\mylenA\usebox\myboxB}{\usebox\myboxA}%
	\else
	\hskip -0.5\mylenA\rlap{\usebox\myboxA}{\hskip 0.5\mylenA\usebox\myboxB}%
	\fi}
\makeatother

\def\comment#1{}

\newcommand{\stkout}[1]{
	\color{red}\ifmmode\text{\sout{\ensuremath{#1}}}\else\sout{#1}\fi\color{black}}

\begin{document}

\title{Blind Modulo Analog-to-Digital Conversion}

\author{Amir Weiss, Everest Huang, Or Ordentlich and Gregory W. Wornell
\thanks{A. Weiss and G. W. Wornell are with the Massachusetts Institute of Technology, Cambridge, MA 02139, USA (email: \{amirwei,gww\}@mit.edu), E. Huang is with MIT Lincoln Laboratory, Lexington, MA 02421 (email: everest@ll.mit.edu),  O. Ordentlich is with the Hebrew University of Jerusalem, Jerusalem 91904, Israel (email: or.ordentlich@mail.huji.ac.il).}
\thanks{
This material is based upon work supported, in part, by the United States Air Force under Air Force Contract No.~FA8702-15-D-0001. Any opinions, findings, conclusions or recommendations expressed in this material are those of the author(s) and do not necessarily reflect the views of the United States Air Force. This work was also supported, in part, by ISF under Grant 1791/17, and NSF under Grant CCF-1717610.}
\thanks{DISTRIBUTION STATEMENT A. Approved for public release. Distribution is unlimited. \textcopyright\, 2021 Massachusetts Institute of Technology.
}
\thanks{Delivered to the U.S. Government with Unlimited Rights, as defined in DFARS Part 252.227-7013 or 7014 (Feb 2014). Notwithstanding any copyright notice, U.S.\ Government rights in this work are defined by DFARS 252.227-7013 or DFARS 252.227-7014 as detailed above. Use of this work other than as specifically authorized by the U.S.\ Government may violate any copyrights that exist in this work.}
}
	%
		
		

\maketitle

\begin{abstract}
In a growing number of applications, there is a need to digitize signals whose spectral characteristics are challenging for traditional Analog-to-Digital Converters (ADCs). Examples, among others, include systems where the ADC must acquire at once a very wide but sparsely and dynamically occupied bandwidth supporting diverse services, as well as systems where the signal of interest is subject to strong narrowband
co-channel interference. In such scenarios, the resolution requirements can be prohibitively high. As an alternative, the recently proposed \emph{modulo-ADC} architecture can in principle require dramatically fewer bits in the conversation to obtain the target fidelity, but requires that information about the spectrum be known
and explicitly taken into account by the analog and digital processing in the converter, which is frequently impractical. To address this limitation, we develop a \emph{blind} version of the architecture that requires no such knowledge in the converter, without sacrificing performance. In particular, it features an automatic modulo-level adjustment and a fully adaptive modulo unwrapping mechanism, allowing it to asymptotically match the characteristics of the unknown input signal. In addition to detailed analysis, simulations demonstrate the attractive performance characteristics in representative settings.

\end{abstract}


\section{Introduction}\label{sec:intro}
The available spectrum for a communication system is increasingly congested and varies widely by location and time. One strategy for operation in these dynamic conditions is to scan the spectrum to find unoccupied bandwidth within which to transmit. Historically, this has been a difficult task since the fraction of all the potentially usable bandwidth that can be scanned simultaneously is limited by the bandwidth of the receiver front-end. In a traditional system architecture, a narrow analog filter matched to the desired communication band rejects out-of-band transmissions from overwhelming the Analog-to-Digital Converter (ADC) prior to any digital processing.  This results in either a fixed frequency communication system, or an expensive frequency-agile analog front-end for both the transmitter and receiver. For a narrowband system, the fraction of potentially available bandwidth that can be monitored at any instant can be small, which slows down the response to potentially rapidly changing channel conditions.  The system is further complicated by the need to coordinate between communication nodes what frequencies are being used at what time.

\begin{figure}[t]
	\includegraphics[width=0.5\textwidth]{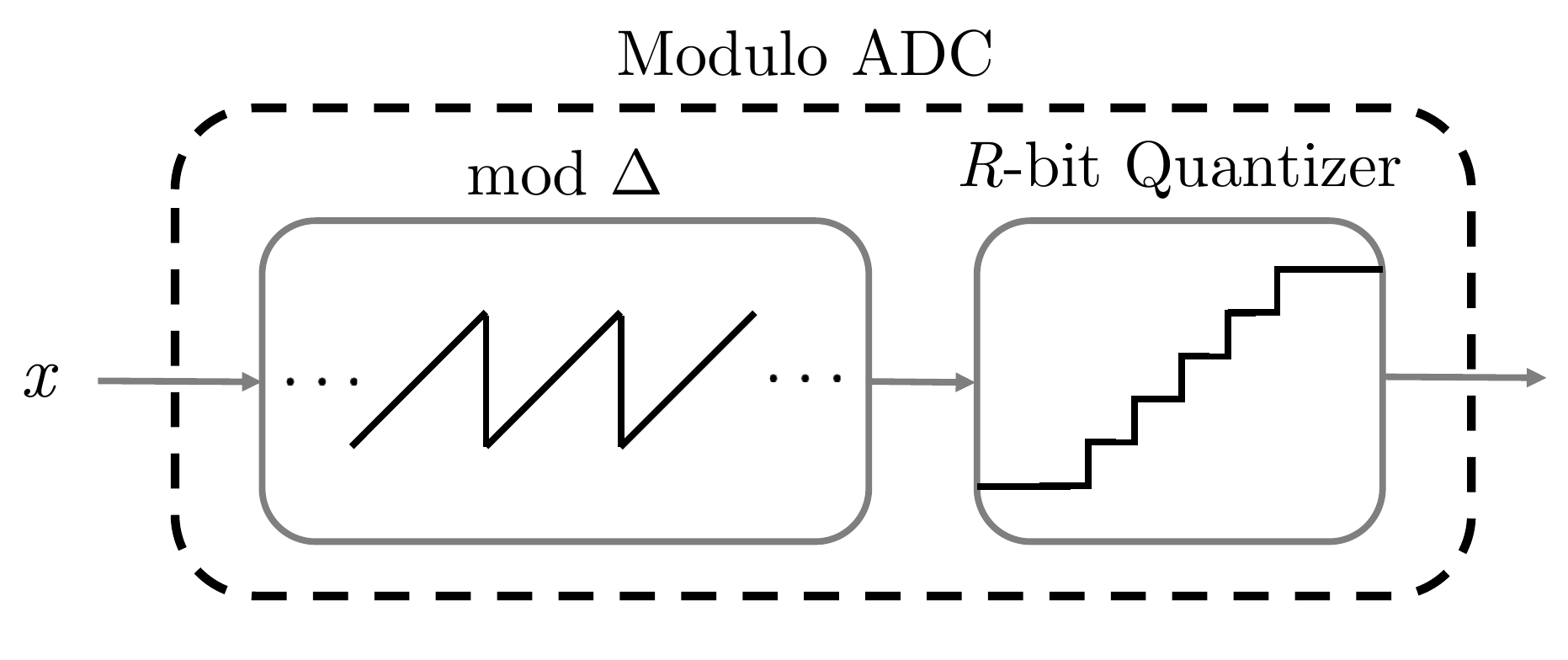}
	\centering
	\caption{A schematic block diagram illustration of the mod-ADC.}
	\label{fig:ModADC_diag_block}\vspace{-0.2cm}
\end{figure}

The emergence of high speed ADCs with multiple GHz of bandwidth enables affordable systems to be built, that can simultaneously scan large regions of the spectrum for unutilized bandwidth to transmit in \cite{wang2010advances}. The congested nature of the spectrum, however, requires robust front-end processing to accommodate the large dynamic range required from multiple possibly strong interfering sources \cite{cabric2005physical}. A wide or changing frequency allocation will by necessity allow these outside sources to be sampled as well. While digital processing can in principle remove the effect of the undesired signals, the ADC must still be able to faithfully sample the entire bandwidth, containing all signals, prior to any subsequent digital manipulation. Thus, despite the strong structure and high predictability of the sampled signal, a traditional ADC requires a large number of bits per-second in order to allow for high-quality reconstruction. 

\begin{figure*}
	\centering
	\begin{subfigure}{0.48\textwidth}
		\centering
		\includegraphics[width=\textwidth]{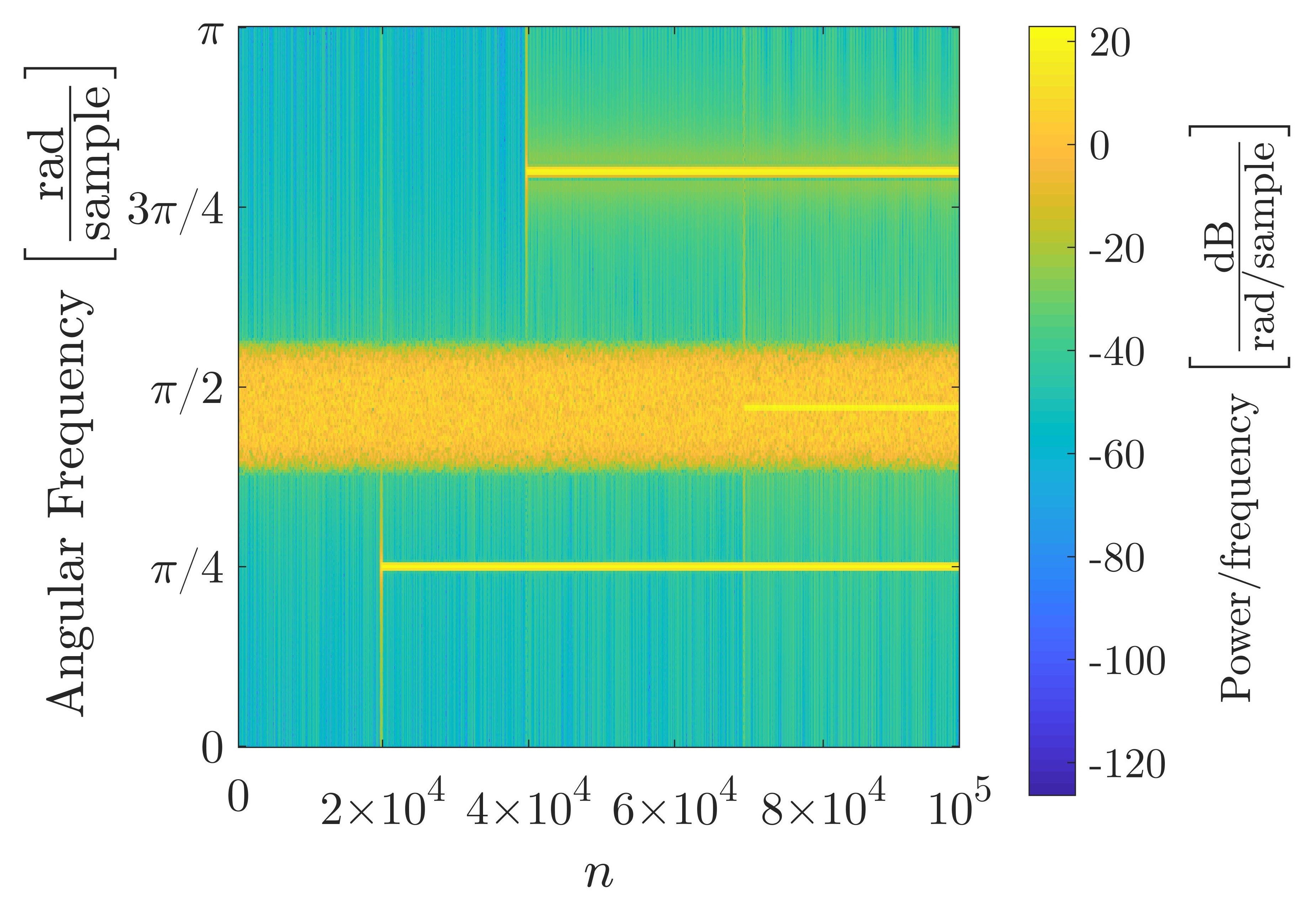}\vspace{-0.1cm}
		\caption{}
		\label{fig:introfigurea}
	\end{subfigure}
	\begin{subfigure}{0.48\textwidth}
		\centering
		\includegraphics[width=\textwidth]{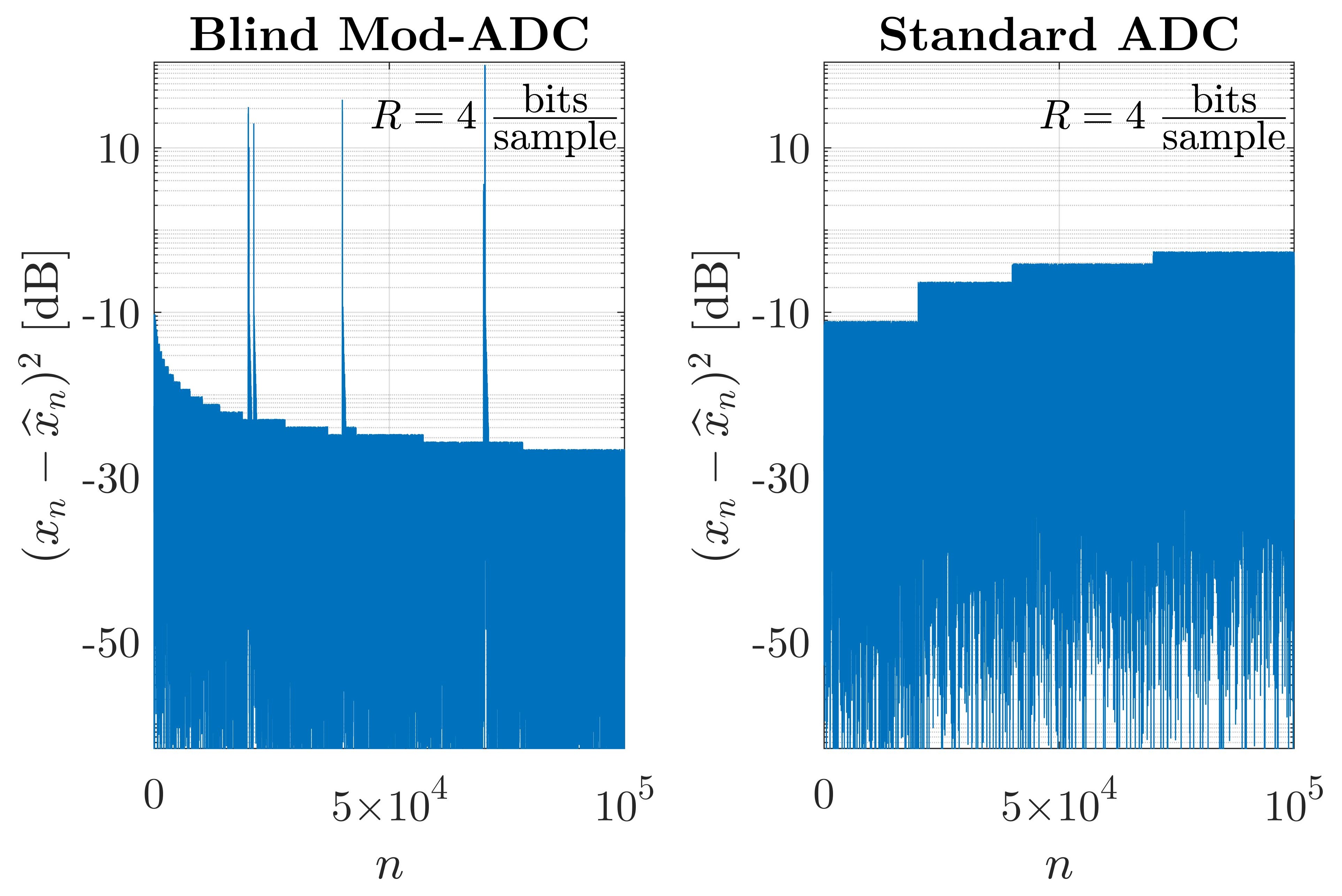}\vspace{-0.1cm}
		\caption{}
		\label{fig:introfigurec}
	\end{subfigure}
	\begin{subfigure}{0.96\textwidth}
		\centering
		\includegraphics[width=\textwidth]{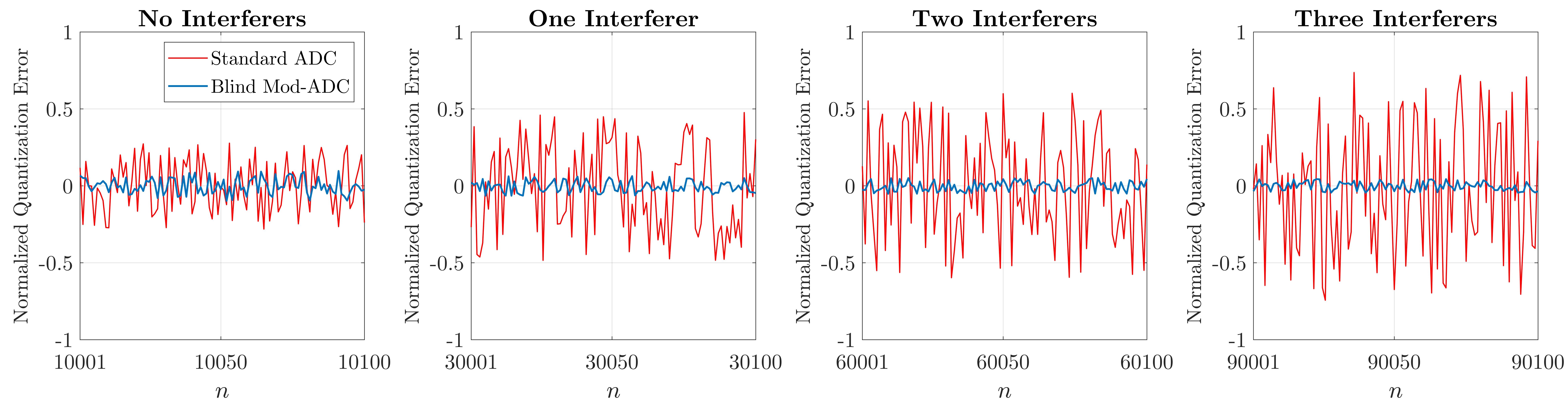}\vspace{-0.1cm}
		\caption{}
		\label{fig:introfigureb}
	\end{subfigure}
	\caption{An illustrative scenario of interest. (a) Spectrogram of the ADC input: a bandlimited signal of interest in the presence of narrowband interferences (b) The instantaneous squared errors of the outputs of a standard ADC and our proposed blind mod-ADC (c) Four snippets of the time-domain normalized quantization errors of the two ADCs. Evidently, the blind mod-ADC achieves a significantly lower MSE relative to a standard uniform ADC with a perfect (``oracle") AGC.}
	\label{fig:introfigures}
\end{figure*}

One possible approach for addressing the inefficiency described above, is using \emph{modulo ADCs}~\cite{ordentlich2018modulo} instead of a traditional ADC. A modulo ADC first folds each sample of the input process modulo $\Delta$, where $\Delta$ is a design parameter, and only then quantizes the result using a traditional uniform quantizer. See Fig.\ \ref{fig:ModADC_diag_block} for a schematic description. The modulo operation limits the dynamic range of the signal to be quantized, which in turns results in a quantization error whose magnitude is proportional to $\Delta$, rather than to the dynamic range of the original, unfolded signal. In~\cite{ordentlich2018modulo} it is shown that the obtained signal can be reliably unfolded, provided that $\Delta$ is appropriately chosen proportionally to the standard deviation of the \emph{prediction error} in predicting the (quantized) input from its past. Thus, when using a modulo ADC for digitizing a highly predictable process, one can attain high resolution using far fewer bits than for a white process. Simple recovery algorithms based on linear prediction are also given in~\cite{ordentlich2018modulo}.

A major caveat of the modulo ADC framework developed in~\cite{ordentlich2018modulo} is that it assumes knowledge of the Second-Order Statistics (SOSs) of the input process. Such knowledge is crucial for optimizing the modulo ADC parameter $\Delta$, as well as for optimizing the coefficients of the prediction filter used in the unwrapping process. This is a significant impediment to the implementation of modulo ADCs in practice, as commercial ADCs must be robust to the characteristics of the process to be digitized. In particular, traditional ADCs employ an Automatic Gain Control (AGC) mechanism \cite{sun2010automatic} for adapting the quantizers dynamic range to that of the input signal.


In this work we develop a \emph{blind} mechanism for modulo ADCs, which adapts the effective modulo size (analogously to an AGC mechanism in a standard ADC, see Subsection~\ref{subsec:motivationandcontributions}), as well as the coefficients of the unwrapping algorithm prediction filter, to the unknown statistics of the input signal, resulting in a robust ADC architecture. For a stationary input process, the performance of the developed  architecture converges to that of the ``informed'' architecture in~\cite{ordentlich2018modulo}. The developed architecture is also dynamic, and quickly adapts to changes in the characteristics of the input signal.


\subsection{A Motivating Example}\label{subsec:example}

To illustrate the challenges in digitizing communication signals whose locations within the frequency band are unknown, we consider the signal whose spectrogram is depicted in Fig.\ \ref{fig:introfigurea}. This spectrogram corresponds to a sampled Binary Phase-Shift Keying (BPSK) signal together with three narrowband interfering signals (specifically, pure tones), with each interferer initiated at a different time. If the carrier frequency of the BPSK signal were known in advance, one could first down-convert it and use an analog low-pass filter to cancel out all interference outside the frequency band it occupies, and only then sample at the corresponding Nyquist rate. The discrete-time signal resulting from this process is essentially ``white" (i.e., temporally uncorrelated), at least before the interfering tone is initiated, and a standard uniform ADC would efficiently convert it to a sequence of bits.

Unfortunately, as described earlier, estimating the carrier frequency of the communication signal of interest is often a highly challenging task under the required latency constraints. Consequently, the ADC must be applied on the sampled signal depicted in Fig.\ \ref{fig:introfigurea}. A standard ADC is extremely inefficient for such a signal, as it fails to exploit its sparsity in the frequency domain. The Mean-Squared Error (MSE) attained by a standard ADC is dictated primarily by its dynamic range, which is determined via an AGC mechanism. In order to prevent overload errors that result in saturation of the ADC, the dynamic range is set proportionally to the signal's average power. The signal's power is not affected by the fact that the sampling frequency is significantly higher than the size of the essential support of the signal in the frequency domain. Thus, the number of bits per second a standard ADC must output in order to reach some target MSE is significantly increased due to the uncertainty in the carrier frequency of the signal of interest.

To tackle this shortcoming of standard ADCs, this work develops a robust architecture, based on the emerging modulo ADC framework, which efficiently exploits the underlying structure of the acquired signal. Previous work~\cite{ordentlich2018modulo} has shown that when the statistics of the signal to be acquired are known, or when its Power Spectral Density (PSD) is at least confined to a particular frequency interval known in advance, modulo ADCs attain significant performance gains over standard uniform ADCs. Here, we develop an ``AGC equivalent" mechanism for modulo ADCs, that prevents the need for prior knowledge of the signal's statistics; See Fig.\ \ref{fig:Blind_mod_ADC_block_diagram}. This mechanism, together with suitably designed adaptive filtering, results in a \emph{blind} modulo ADC architecture, whose performance approaches that reported in~\cite{ordentlich2018modulo} for stationary signals with known PSD.

Fig.\ \ref{fig:introfigurec} depicts the instantaneous squared error attained by the developed blind modulo ADC architecture for the signal from Fig.\ \ref{fig:introfigurea}. For comparison, we also plot the squared error attained by a standard uniform ADC for the same signal. It is assumed that a perfect AGC is used for the standard ADC, such that its dynamic range is equal to $\kappa\sqrt{\mathrm{Var}(x_n)}$, where $\{x_n\}$ is the input signal, and $\kappa$ is a confidence parameter determining the overload probability. The same value of $\kappa$ is used for both the standard and the modulo ADC systems, such that the overload probabilities for the two systems are similar.\footnote{The overload event for a modulo ADC is the event that the prediction error's magnitude exceeds the dynamic range, as will be explained in detail, and explicitly defined in the sequel.} Furthermore, both the standard and the blind modulo ADC systems use $R=4$ bits per sample.
It is evident that: (i) The developed blind modulo ADC architecture attains a significantly smaller MSE than the one attained by a standard ADC; (ii) It quickly adapts to changes in the characteristics of the input signals, as reflected by Fig.\ \ref{fig:introfigurec}; and (iii) While the addition of narrowband interferers strongly degrades the performance of the standard ADC, the MSE attained by the blind modulo ADC architecture is largely unaffected.

\subsection{Related Work}\label{subsec:relatedwork}

The idea of using modulo ADCs/quantizers for exploiting temporal correlations within a stationary input process towards reducing the quantization rate $R$, dates back, at least, to~\cite{er79}, where a quantization scheme, called modulo-PCM, was introduced. Under the so called ``high-resolution'' assumption, which restricts the quantization's error PSD to be much smaller than that of the signal \emph{for all} frequencies~\cite{zamir2008achieving}, the analysis in~\cite{er79} has shown that this scheme can attain distortion almost as small as the fundamental information theoretic lower bounds. Unfortunately, the ``high-resolution'' assumption breaks down completely for processes whose PSD function is not supported on the entire spectrum. Such processes include, for example, the process from Fig.\ \ref{fig:introfigurea}, as well as any oversampled process. To that end, building on~\cite{zamir2008achieving}, a different modulo unwrapping algorithm was developed in~\cite{ordentlich2018modulo}, and the resulting modulo ADC system was shown to attain distortion close to the fundamental information theoretic lower bounds, even for processes for which the ``high-resolution'' assumption fails. Furthermore, relying on the unwrapping techniques developed in~\cite{oe17}, a modulo ADC framework accompanied by an unwrapping algorithm was developed for vector processes, that are correlated in both space and time. Finally~\cite{ordentlich2018modulo} also developed an architecture for a ring-oscillators-based circuit implementing a modulo ADC.

It should be noted that the results mentioned above rely on complete knowledge of the statistical law governing the inputs to the ADCs, with the exception of the result in~\cite[Section III]{ordentlich2018modulo}, which is robust, but is of a minimax nature, in contrast to the pointwise optimality we seek here. For the case of temporally uncorrelated vector processes, Romanov \emph{et al.}~\cite{romanov2021blind} developed a blind unwrapping algorithm which achieves performance close to that of an informed unwrapping algorithm, fully aware of the statistics. While a stationary process in time can be treated as a vector process in high dimensions, the scaling of the sample complexity of the algorithm from~\cite{romanov2021blind} renders it prohibitive for the blind modulo ADC problem of time processes under consideration in this work.

The line of research described above considers the improvement modulo ADCs offer over standard ADCs in terms of the trade-off between quantization rate and MSE distortion. The current paper continues this line of work. Another line of work which has received attention recently is that of the so-called ``unlimited sampling''. Under the unlimited sampling framework, the quantization noise is usually not accounted for, and the focus is on characterizing the conditions which guarantee that a signal can be reconstructed from its folded version~\cite{bhandari2017unlimited,bkr18isit,bhandari2020unlimited,bk20}. In particular, it was shown that under mild conditions, a continuous time bandlimited signal can be recovered from its modulo reduced samples, provided that the sampling rate exceeds Nyquist's rate~\cite{bhandari2020unlimited,romanov2019above,bk19}, regardless of the modulo size. Some of the more recent work on unlimited sampling~\cite{gbk19,bkp21} does take quantization noise into account, but adopts a worst-case model for the input signal (over a predefined class of signals), whereas here we model the input signal to the ADC as a stochastic process, and accordingly, analyze the statistical behaviour of the MSE.

\subsection{Contributions}\label{subsec:motivationandcontributions}
{
	\setlength{\arrayrulewidth}{0.5mm}
	\setlength{\tabcolsep}{18pt}
	\renewcommand{\arraystretch}{1.5}
\begin{table*}[t]
	\centering
	\begin{tabular}{ |p{7.7cm}|p{7.7cm}|  }
		\hline
		\multicolumn{1}{|c|}{\underline{\textbf{Standard ADC}}} & \multicolumn{1}{c|}{\underline{\textbf{Informed Mod-ADC}}} \\
		\emph{Encoder} side information: Input signal variance, $\sigma_x^2$ & \emph{Encoder} side information: Innovation variance \\
		\emph{Decoder} side information: Input signal variance, $\sigma_x^2$ & \emph{Decoder} side information: PSD of the input signal \\
		\emph{Performance}: Near minimax optimal with variance constraint $\sigma_x^2$ & \emph{Performance}: Near point-wise optimal \\
		\hline
		\multicolumn{1}{|c|}{\underline{\textbf{Standard ADC with AGC}}} & \multicolumn{1}{c|}{\underline{\textbf{Blind Mod-ADC}}} \\
		\emph{Encoder} side information: None & \emph{Encoder} side information: None   \\
		\emph{Decoder} side information: None & \emph{Decoder} side information: None \\
		\emph{Performance}: Near minimax optimal for the (unknown) variance $\sigma_x^2$  & \emph{Performance}: Near point-wise optimal \\
		\hline
	\end{tabular}
	\caption{Summary of the side information available to each type of ADC and the respective performance. The above hold under the assumption that the input signal $\{x_n\}$ to the ADC is a stationary Gaussian process.}
	\label{tab:comparisontableofADCs}
\end{table*}
}
In light of all the above, it is clear that a significant step towards realizing the modulo ADC technology is by developing the algorithmic framework, which will provide the essential robustness with respect to different types of signals and dynamic environments. Hence our motivation is developing an architecture with the appropriate algorithmic framework, which on one hand will be able to adapt quickly to changes reflected in the temporal structure of the input signal, and on the other hand will still provide reliable and stable high-resolution analog-to-digital conversion.

In order to appreciate our contributions, it is instructive to consider the trade-offs exhibited by several ADC architectures, as summarized in Table~\ref{tab:comparisontableofADCs}. The table compares between a standard ADC, a standard ADC with AGC, an informed modulo ADC as described in~\cite{ordentlich2018modulo}, and the blind modulo ADC architecture we develop here. We compare the four solutions in terms of the statistical knowledge they require, and their performance guarantees. To simplify the exposition, suppose, for example, that the input to the ADC $\{x_n\}$ is a zero-mean stationary Gaussian process, with a (possibly unknown) PSD.

A standard (uniform) ADC has a fixed dynamic range. In order for overload events to be rare, such that the ADC is usually not saturated, the dynamic range must be greater than the standard deviation of $\{x_n\}$, denoted by $\sigma_x$, by some constant factor. Thus, in the design of the encoder and decoder, it is implicitly assumed that (an upper bound on) $\sigma_x$ is known. The standard ADC cannot exploit any ``memory'' in the process $\{x_n\}$, but for an i.i.d. process it attains a rate-distortion trade-off which is quite close to the fundamental information theoretic limits, characterized by the rate-distortion function of the source \cite{berger1971ratedistortion}. Thus, it is near minimax optimal with respect to the class of all PSDs with variance $\sigma^2_x$. A standard ADC with an AGC automatically adapts its dynamic range to $\sigma_x$, and does not require prior knowledge of it. Consequently, it attains near minimax optimality for the class of all PSDs with variance $\sigma^2_x$, simultaneously for all values of $\sigma^2_x$.

The informed modulo ADC from~\cite{ordentlich2018modulo} requires the encoder to set the modulo size (or the signal scaling) appropriately, which requires knowledge of the variance of the innovation process (i.e., the error process due to optimal prediction). The decoder requires knowledge of the entire PSD in order to compute the coefficients of the optimal prediction filter it uses. It was shown~\cite{ordentlich2018modulo} that for input processes of finite differential entropy rate, the rate-distortion trade-off this architecture attains is near optimal, as the quantization rate increases.

Clearly, a commercial ADC cannot be designed under the assumption that the innovation variance and the entire PSD of the input process is known in advance. In this paper, we close this gap and develop the blind modulo ADC architecture that makes no assumptions on the input process in the design of the encoder and the decoder, but nevertheless attains the same asymptotic performance as the modulo ADC architecture from~\cite{ordentlich2018modulo}. In particular, our developed architecture asymptotically nearly attains the optimal rate-distortion trade-off \emph{simultaneously} for all process with a finite differential entropy rate. The blind modulo ADC scheme we develop here adapts the modulo size / signal scaling at the encoder according to the associated innovation variance of the input process. Note that this task is considerably more challenging than that of an AGC in a standard ADC, since estimating the innovation variance is more involved than estimating the variance itself. Moreover, the decoder in a blind modulo ADC is implicitly estimating the necessary SOSs (for means of optimal prediction) beyond merely variance, i.e., cross correlations between past and present samples. Furthermore, the decoder blindly unwraps the quantized signal from the modulo measurements.

Our two main contributions in this work are the following:
\begin{itemize}
\item \emph{Adaptive Algorithm for Blind Modulo Unwrapping:} We propose a feedback solution algorithm for a modulo ADC encoder-decoder, which blindly unwraps the modulo folding of the input signal. That is, our algorithm does not use prior knowledge on the temporal structure (i.e., the autocorrelation function) of input signal. Nevertheless, using the Least Mean Squares (LMS) algorithm \cite{haykin2003least}, we are able to learn (only) the required SOSs, which allow us to exploit the unknown temporal structure, and gradually increase the resolution of the modulo ADC. Consequently, our developed blind modulo ADC architecture is more robust and practical than the modulo ADC architecture from~\cite{ordentlich2018modulo}, which is designed based on such prior knowledge.
\item \emph{Asymptotic Performance Analysis of the Blind Modulo ADC Architecture:} We analyze the asymptotic performance of the developed algorithm in terms of the attainable resolution. We derive and present an insightful closed-form expression for the MSE distortion, which not only forecasts the best attainable performance under the specified conditions (dictated by the system parameters), but also intuitively explains the fundamental accuracy-stability trade-off inherent to the blind nature of the problem under consideration. Moreover, a steady state detector naturally stems from this analysis, allowing us to estimate the time at which the adaptive process can be (locally) paused. Consequently, the stability of the proposed method is increased, and as a (positive) byproduct, the overall computational load is reduced.
\end{itemize}

\subsection{Paper Organization}\label{subsec:paperorganization}
The rest of the paper is organized as follows. The remainder of this section is devoted to a short outline of our notations. Section \ref{sec:idealmoduloADC} is devoted to a brief review of the modulo ADC framework previously presented in \cite{ordentlich2018modulo}, setting the premises for the current work. In Section \ref{sec:problemformulation} we formulate the problem of blind modulo ADC. Our proposed adaptive solution algorithm is presented in Section \ref{sec:proposedsolution}, where we derive the different algorithmic components in separate subsections, discuss key system parameters, trade-offs, and the asymptotic performance. Simulation results, corroborating our analytical derivation, are presented in Section \ref{sec:simulresults}, and concluding remarks are given in Section \ref{sec:conclusion}.

\subsection{Notations}\label{subsec:notations}
We use $x$ and $\ux$ for a scalar and a column vector, respectively. The superscript $(\cdot)^{\tps}$ denotes the transposition. We use $\mathbbm{1}_{\mathcal{A}}$ to denote the indicator function of the event $\mathcal{A}$, namely $\mathbbm{1}_{\mathcal{A}}=1$ if $\mathcal{A}$ is true, and $\mathbbm{1}_{\mathcal{A}}=0$ otherwise. $\Eset[\cdot]$ and ${\mathrm{Var}}(\cdot)$ denote expectation and variance, respectively, and $\Tr(\cdot)$ denotes the trace operator. We use $\widehat{\;}$ to denote an estimator, e.g., $\widehat{x}$ is an estimator of $x$.

\section{Review on a Modulo ADC}\label{sec:idealmoduloADC}
In this section, we briefly review the modulo ADC (encoding-decoding) algorithm previously proposed in \cite{ordentlich2018modulo} for scalar stationary processes. As our proposed blind method relies on some similar fundamental concepts, it is instructive to review the ``informed" algorithm, which is described below.

For a positive number $\Delta\in\Rset^+$, we define
\begin{equation}\label{modulodefinition}
[x]\;{\rm{mod}}\;\Delta\triangleq x-\Delta\cdot\left\lfloor \frac{x}{\Delta}\right\rfloor\in[0,\Delta), \quad \forall x\in\Rset,
\end{equation}
as the $[\cdot]\;{\rm{mod}}\;\Delta$ operator, where $\left\lfloor x\right\rfloor$ is the floor operation, which returns the largest integer smaller than or equal to $x$. An $R$-bit modulo ADC with resolution parameter $\alpha$, termed $(R,\alpha)$ mod-ADC, produces its output by first computing
\begin{equation}\label{modADCdefinition}
[x]_{R,\alpha}\triangleq\left[\left\lfloor \alpha x\right\rfloor\right]\;{\rm{mod}}\;2^R\in\{0,1,\ldots,2^R-1\},
\end{equation}
and then producing the binary representation of \eqref{modADCdefinition}. A schematic illustration of the mod-ADC is given in Fig.\ \ref{fig:ModADC_diag_block}.

Notice that when writing $[x]_{R,\alpha}$ as 
\begin{equation}\label{quantnoise}
[x]_{R,\alpha}=[\alpha x +\underbrace{\left(\left\lfloor \alpha x\right\rfloor-\alpha x\right)}_{\triangleq \widetilde{z}}]\;{\rm{mod}}\;2^R=\left[\alpha x +\widetilde{z}\right]\;{\rm{mod}}\;2^R,
\end{equation}
we identify $\widetilde{z}\in(-1,0]$ as the quantization error of a uniform scalar quantizer \cite{gray1998quantization}. Although $\widetilde{z}$ is a deterministic function of $x$, this quantization error can be modeled quite accurately as additive \emph{random} uniform noise. For details on the justification of this assumption by using subtractive dithers \cite{lipshitz1992quantization}, see \cite{ordentlich2018modulo}. Under this assumption, an $(R,\alpha)$ mod-ADC is viewed as a stochastic channel, whose output $y$ for an input $x$ is given by
\begin{equation}\label{modulorandomchannel}
y=\left[\alpha x+z\right]\;{\rm{mod}}\;2^R,
\end{equation}
where $z\sim{\rm{Unif}}\left((-1,0]\right)$. Obviously, since the modulo operation is a form of \emph{lossy} compression, it is generally impossible to recover the unfolded signal $\alpha x+z$ from its folded version $y=\left[\alpha x+z\right]\;{\rm{mod}}\;2^R$. Nevertheless, under relatively mild conditions, when the input signal is ``temporally-predictable" to a sufficient degree, e.g., a correlated random process \cite{ordentlich2018modulo} or a deterministic bandlimited signal \cite{bhandari2017unlimited,romanov2019above}, it is in fact possible to \emph{perfectly} recover the unfolded signal\footnote{With high probability (w.h.p.) for random signals, and to an arbitrary precision for deterministic bandlimited signals (``w.h.p." in the sense that the probability of prefect recovery can be made arbitrarily large by increasing $R$).} from its past samples and its current folded sample via causal processing.

More specifically, consider an $(R,\alpha)$ mod-ADC whose input signal $x_n$ is a zero-mean stationary random process, with a known autocorrelation function $R_{x}[\ell]\triangleq\Eset\left[x_nx_{n-\ell}\right]\in\Rset$, whose one-sided support is assumed to be at least of (discrete) length $p\in\Nset^+$. The output of the mod-ADC is given by
\begin{equation}\label{modulooutputprocess}
y_n=\left[\alpha x_n+z_n\right]\;{\rm{mod}}\;2^R,\;\;\forall n\in\Nset^+,
\end{equation}
where $\{z_n\sim {\rm{Unif}}((-1,0])\}$, modeling the quantization noise, is an independent, identically distributed (i.i.d.) stochastic process. Further, define the unfolded quantized signal,
\begin{equation}\label{defofnonfolded}
v_n\triangleq\alpha x_n + z_n,\;\;\forall n\in\Nset^+,
\end{equation}
and assume that the decoder has access to $\{v_{n-1},\ldots,v_{n-p}\}$, which is equivalent to assuming that the last $p$ samples of $y_n$ were correctly decoded. This can be achieved, for example, by proper initialization with a sufficiently small resolution parameter $\alpha$, a notion that will also be used as part of our proposed method. For additional justifications of this assumption, see \cite{ordentlich2018modulo}, Subsection II-A. Note that once $v_n$ is recovered, $x_n$ is readily estimated as $\widehat{x}_n=\frac{v_n+\frac{1}{2}}{\alpha}$. Thus, we focus on recovering $v_n$ from $y_n$ and $\uv_n\triangleq[v_{n-1} \cdots v_{n-p}]^{\tps}\in\Rset^{p\times 1}$.

The decoding algorithm proposed in \cite{ordentlich2018modulo} for recovering $v_n$ w.h.p.\ when $R_x[\ell]$ is \emph{known}, here referred to as \emph{oracle} modulo unfolding, is given in Algorithm \ref{Algorithm1}. The main idea behind the prescribed technical steps is the following. Every number $\gamma\in\Rset^+$ (similarly for $\gamma\in\Rset^-$) can be represented as
\begin{equation}\label{deltarepresentationofgamma}
\gamma=(\underbrace{\Delta+\Delta+\ldots+\Delta}_{K_{\gamma}\in\Nset^+ \text{ times}}) + \underbrace{[\gamma]\;{\rm{mod}}\;\Delta}_{\triangleq \varepsilon_{\gamma}}=K_{\gamma}\Delta+\varepsilon_{\gamma},
\end{equation}
where, intuitively, $K_{\gamma}$ and $\varepsilon_{\gamma}$ correspond to coarse and fine information, respectively, in the ``$\Delta$-representation" \eqref{deltarepresentationofgamma}. The mod-ADC records only the fine information $\varepsilon_{\gamma}$ in $\gamma$. Hence, for perfect reconstruction, only $K_{\gamma}$ is required (assuming $\Delta$ is known). Conceptually, this means that as long as an estimator of $v_n$ (possibly linear) has a minimal accuracy level, such that its residual estimation error lie in $[0,\Delta)$, $K_{\gamma}$ can be recovered, which, in turn, means that $v_n$ can be perfectly recovered.
\begin{algorithm}[t]
	\KwIn{$y_n, \uv_n, R_x[\ell], \alpha, R$}
	\KwOut{$\widehat{v}_{\text{oracle},n}$}
	\nl Compute the Linear Minimum Mean Squared Error (LMMSE) estimate of $v_n$ based on $\uv_n$
	\begin{equation}\label{LMMSElengthp}
	\widehat{v}_{\text{\tiny LMMSE},n}^p=\uh_{\text{opt}}^{\tps}\left(\uv_n+\frac{1}{2}\uone\right)-\frac{1}{2},
	\end{equation}
	where $\uh_{\text{opt}}\in\Rset^{p\times 1}$ is the length-$p$ Finite Impulse Response (FIR) filter yielding the LMMSE estimator, computed based on $R_x[\ell]$, and the shifts are to compensate for the $\Eset[z_n]=-\frac{1}{2}$;\\
	\nl Compute
	\begin{align}
	w_{\text{\tiny LMMSE},n} &= [y_n-\widehat{v}_{\text{\tiny LMMSE},n}^p]\;{\rm{mod}}\;2^R,\\
	{\widehat{e}}_{\text{\tiny LMMSE},n}^{\,p} &\triangleq \left(\left[w_{\text{\tiny LMMSE},n}+\frac{1}{2}2^R\right]\;{\rm{mod}}\,2^R\right)-\frac{1}{2}2^R;
	\end{align}\\
	\nl Return $\widehat{v}_{\text{oracle},n}=\widehat{v}_{\text{\tiny LMMSE},n}^p+\widehat{e}_{\text{\tiny LMMSE},n}^{\,p}$.
	\caption{{\bf Oracle Modulo Unfolding} \newline $\widehat{v}_{\text{oracle},n}=\text{ModUnfold}(y_n, \uv_n, R_x[\ell], \alpha, R)$ \label{Algorithm1}}
\end{algorithm}
\setlength{\textfloatsep}{2pt}

An elaborate analysis of Algorithm \ref{Algorithm1} is provided in \cite{ordentlich2018modulo}, wherein analytical performance guarantees are derived in the form of upper bounds on the probability of the \emph{overload} event, which inflicts $\widehat{v}_n\neq v_n$, and is defined as
\begin{equation}\label{nooverloadeventdefinition}
\mathcal{E}^*_{\tiny {\rm{OL}}_n}\triangleq\left\{\left|e_{\text{\tiny LMMSE},n}^p\right|\geq\frac{1}{2}2^R\right\}=\left\{\widehat{e}_{\text{\tiny LMMSE},n}^{\,p} \neq e_{\text{\tiny LMMSE},n}^p \right\},
\end{equation}
where $e_{\text{\tiny LMMSE},n}^p\triangleq v_n-\widehat{v}_{\text{\tiny LMMSE},n}^p$, and on the conditional Mean Squared Error (MSE) distortion,
\begin{equation}\label{MSEdistortion}
D\triangleq\Eset\left[(x_n-\widehat{x}_n)^2|\mathcal{E}^*_{\tiny \mybar{\rm{OL}}_n}\right].
\end{equation}
Specifically, it was shown that (Proposition 1, \cite{ordentlich2018modulo}),
\begin{align}
\Pr\left(\mathcal{E}^*_{\tiny {\rm{OL}}_n}\right)&\leq 2\exp\left\{-\frac{3}{2}2^{2\left(R-\frac{1}{2}\log_2(12\sigma_{\text{\tiny LMMSE},p}^2)\right)}\right\}, \label{upperbound1}\\
D &\leq \frac{1}{12\alpha^2\left(1-\Pr\left(\mathcal{E}^*_{\tiny {\rm{OL}}_n}\right)\right)},\label{upperbound2}
\end{align}
where $\sigma_{\text{\tiny LMMSE},p}^2\triangleq\Eset\left[\left(e_{\text{\tiny LMMSE},n}^p\right)^2\right]$ is the MSE of the LMMSE estimator based on the previous $p$ samples, as in \eqref{LMMSElengthp}. 

Algorithm \ref{Algorithm1}, along with its information-theoretic analysis \cite{ordentlich2018modulo}, provide strong evidence regarding the potential feasibility and merits of mod-ADCs, which are attractive for approaching the minimal number of raw output bits per sample, for a given sampling frequency $f_s$ and a prespecified distortion level $D$. 

Yet, devices such as ADCs usually operate under dynamic conditions, giving rise to a wide range of possible inputs with unknown characteristics, and must still maintain proper operation. Therefore, one significant step towards implementing mod-ADCs for real-life applications can be made by relaxing the (sometimes too restrictive) assumption that $R_x[\ell]$ is known. We take this significant step in the next sections.

\section{Problem Formulation}\label{sec:problemformulation}
Consider an $(R,\alpha_n)$ mod-ADC as described in the previous section, with a fixed modulo range $\Delta=2^R$, but an \emph{adaptable}, possibly \emph{time-varying} resolution parameter $\alpha_n\in\Rset^+$. The mod-ADC is fed with the input discrete-time signal $\{x_n\triangleq x(nT_s)\}_{n\in \Nset^+}$, acquired by sampling the analog, continuous-time signal $x(t)$ every $T_s=f_s^{-1}$ seconds. We assume that $x_n$ is a zero-mean stationary stochastic process with an \emph{unknown} autocorrelation function $R_{x}[\ell]\triangleq\Eset\left[x_nx_{n-\ell}\right]\in\Rset$. The observed, distorted signal at the output of the mod-ADC reads
\begin{equation}\label{outputofmodADC}
y_n=[\alpha_n x_n+z_n]\;{\rm{mod}}\;2^R, \; \forall n\in\Nset^+,
\end{equation}
where, as before, the quantization noise process $\{z_n\sim {\rm{Unif}}((-1,0])\}$ is i.i.d. Further, we redefine the unfolded quantized signal,
\begin{equation}\label{defofv}
v_n\triangleq\alpha_n x_n + z_n, \; \forall n\in\Nset^+,
\end{equation}
which, in general, is no longer stationary. Nonetheless, when $\alpha_n$ is held fixed on a specific time interval, then $v_n$ can be regarded as stationary on that particular interval.
\begin{figure*}[t]
	\includegraphics[width=0.8\textwidth]{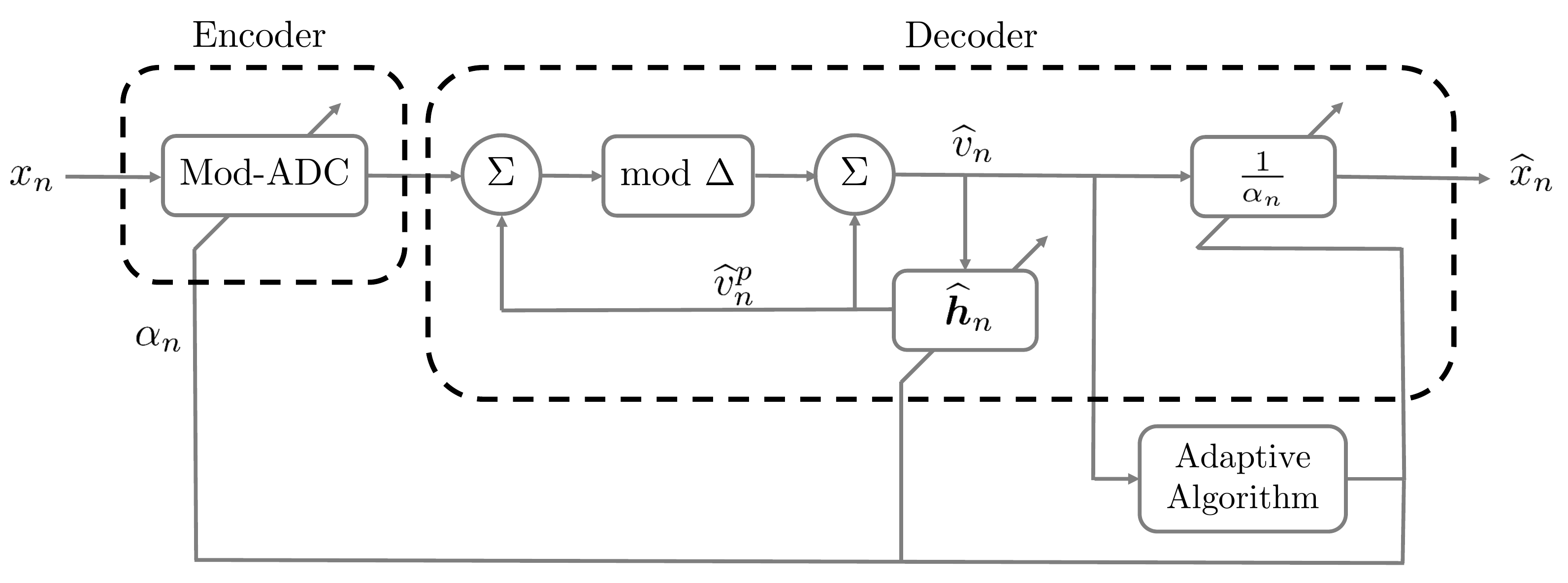}
	\centering
	\caption{A schematic block diagram of the blind mod-ADC encoder-decoder. A block with a diagonal arrow represents an adaptive operation (e.g., filtering).}
	\label{fig:Blind_mod_ADC_block_diagram}\vspace{-0.2cm}
\end{figure*}

The primary goal in this context is to estimate the input signal $x_n$ as accurately as possible based on the observed sequence $\{y_n\}$ at the output of the mod-ADC using a causal system. However, since $v_n$ is merely a scaled version of $x_n$ contaminated by white noise \eqref{defofv}, the problem essentially boils down to recovering $v_n$, and is stated concisely as follows.\vspace{0.2cm}
\textbf{Problem Statement:} {\myfontb For a fixed number of bits $R$, design an adaptive mechanism for estimating $\{x_n\}$ from the output of the mod-ADC with the lowest possible MSE distortion, \emph{without prior knowledge on $R_x[\ell]$}.
}\vspace{0.2cm}

\noindent An interpretation of this problem statement is to design an update mechanism for maximizing the resolution parameter $\alpha_n$, while still allowing for reliable recovery of $v_n$ from $\left\{y_k\right\}_{k\leq n}$, and design the recovery mechanism.

As explained in Section \ref{sec:idealmoduloADC}, provided $v_n$ is \emph{exactly} recovered w.h.p., i.e., $\widehat{v}_n\overset{\text{w.h.p.}}{=}v_n$, the input signal is readily estimated as
\begin{equation}\label{estimateofx}
\widehat{x}_n\triangleq\frac{\widehat{v}_n+\frac{1}{2}}{\alpha_n},
\end{equation}
where $\alpha_n$ is a known system parameter, and $\frac{1}{2}$ is to compensate for the quantization noise (non-zero) expectation $\Eset[z_n]=-\frac{1}{2}$.

\section{Blind Modulo ADC Conversion}\label{sec:proposedsolution}
In this section, we present the blind mod-ADC algorithm, which simultaneously estimates the input signal $x_n$ and performs online learning of the (possibly time-varying) SOSs of the unfolded quantized signal \eqref{defofv}, necessary for estimation of $x_n$. We note that a key characterizing quantity of interest, to be used at some parts throughout the derivation which follows, is the ratio
\begin{equation}\label{effectivemodulorange}
M_n\triangleq\frac{\Delta}{\alpha_n}=\frac{2^R}{\alpha_n}\in\Rset^+,
\end{equation}
dubbed the \emph{effective modulo range}, rather than $\Delta$ or $\alpha_n$ individually. Although theoretically $M_n$ could be adapted by fixing the resolution parameter and adapting the modulo range, due to practical considerations in the actual implementation of the modulo operation, we keep $\Delta$ fixed, and vary the resolution parameter $\alpha_n$. This mechanism can be realized by changing the gain of the input $x_n$ before feeding it to the mod-ADC.

The structure of the proposed blind mod-ADC is depicted in Fig.\ \ref{fig:Blind_mod_ADC_block_diagram}. Note that, in contrast to an informed mod-ADC (\textit{cf}.\ Fig.\ 3 in \cite{ordentlich2018modulo}), here both the encoder and decoder are adaptive, and vary with time according to the statistical properties of the input signal. The price paid for the expected robustness we enjoy by using the blind mod-ADC is mainly in the form of an \emph{adaptive} filter, rather than a pre-defined, constant one.

The underlying concept of our approach is the following. For a fixed resolution parameter $\alpha_n$, given that at any time instance $n$ the unfolded signal $v_n$ can be \emph{exactly} recovered, we may estimate the optimal length-$p$ FIR filter $\uh_{\text{opt}}$, corresponding to the optimal LMMSE estimator of $v_n$ based on the last consecutive $p$ samples $\{v_{n-1},\ldots,v_{n-p}\}$. This can be done, e.g., using the celebrated LMS algorithm \cite{haykin2003least}, which converges\footnote{In the mean sense, under mild conditions stated explicitly in the sequel.} to $\uh_{\text{opt}}$. Upon convergence, the resolution parameter $\alpha_n$ can be slightly increased, and as long as the estimation error of the linear causal estimator---currently no longer optimal---is sufficiently small, $v_n$ could still be recovered using the same technique as in Algorithm \ref{Algorithm1}. Now, fixing $\alpha_n$ again to its new value, the FIR filter can be adapted again to the optimal one using the LMS algorithm. The process is repeated until a certain level of effective modulo range is attained. This level, reflecting the desired trade-off between the MSE distortion $D$ \eqref{MSEdistortion} and the probability of an overload event $\mathcal{E}^*_{\tiny {\rm{OL}}_n}$ \eqref{nooverloadeventdefinition}, will be later on discussed in detail.

Intuitively, and informally, only appropriate initial conditions and sufficiently smooth transitions from one resolution level to another are required for convergence of the above adaptive process. Conceptually, once these are fulfilled, we attain successful steady state operation of a blind mod-ADC (i.e., $R_x[\ell]$ unknown), in the desired effective modulo range.

Fortunately, with careful attention to more, important and relevant, details, this idea can be realized, and is rigorously described as our algorithm in the following subsections.

\subsection{Phase 1: Initialization}\label{subsec:phase1initialization}
We begin with a ``small" initial value for the resolution parameter, $\alpha_0$ (equivalently, $M_0=2^R/\alpha_0$), that ensures an essentially degenerated modulo operation, i.e., $\widetilde{y}_n=v_n$, where
\begin{equation}\label{degeneratemodulo2}
\widetilde{y}_n\triangleq\left(\left[y_n+\frac{1}{2}2^R\right]\;{\rm{mod}}\;2^R\right)-\frac{1}{2}2^R,
\end{equation}
such that $\widetilde{y}_n$ is the ``modulo-shifted" version of $y_n$. Note that, since $x_n$ is zero-mean, $y_n$ actually undergoes a modulo operation quite often (roughly half of the time when $\Pr(x_n<0)=0.5$). However, this is not an essential modulo due to a large amplitude of the input $\alpha_nx_n$, and is merely due to the fact that the dynamic range under consideration is $[0,2^R)$, rather than $[-\frac{1}{2}2^R,\frac{1}{2}2^R)$. Nevertheless, we stick to this formulation as it more accurately describes the actual realization of our proposed method. For purposes that will become clear in the sequel, we further define for convenience
\begin{equation}\label{normalizedv}
\mybar{v}_n\triangleq\frac{v_n+\frac{1}{2}}{\alpha_n}= x_n + \frac{z_n+\frac{1}{2}}{\alpha_n},
\end{equation}
the ``$\alpha_n$-standardized" version of $v_n$. Note that \eqref{normalizedv} still depends on the resolution parameter $\alpha_n$. However, since
\begin{align}
\Eset\left[\mybar{v}_n\right]&=\Eset\left[x_n\right]+\frac{\Eset\left[z_n\right]+\frac{1}{2}}{\alpha_n}=0,\label{meanofvtilde}\\
\Eset\left[\mybar{v}_n^2\right]&=\Eset\left[x_n^2\right]+\frac{\Eset\left[\left(z_n+\frac{1}{2}\right)^2\right]}{\alpha_n^2}\triangleq\sigma_x^2+\frac{1}{12\alpha_n^2},\label{powerofvtilde}
\end{align}
when $\alpha_n$ is sufficiently large, the variance of $\mybar{v}_n$ is dominated by $\sigma_x^2$, and can be considered to be approximately independent of the system parameter $\alpha_n$ for certain needs. Of course, during the initialization phase, this is (still) not the case.

Assuming that $\widetilde{y}_n=v_n$ as long as $\alpha_n=\alpha_0$ is fixed, the optimal length-$p$ FIR filter for estimation of $v_n$ \eqref{defofv} based on $\ubv_n\triangleq[\mybar{v}_{n-1}\cdots\mybar{v}_{n-p}]^{\tps}\in\Rset^{p\times1}$ \eqref{normalizedv}, which is merely a shifted-scaled version of $\uv_n$, can be estimated with the LMS algorithm using the following update equation \cite{haykin2003least},
\begin{equation}\label{LMSequation}
\widehat{\uh}_n=\widehat{\uh}_{n-1}+\mu\cdot\ubv_ne_n^p.
\end{equation}
Here, $\widehat{\uh}_n$ is the FIR filter used in Algorithm \ref{Algorithm2} for the recovery of $v_n$, $\mu$ is the learning rate (or step size), and 
\begin{equation}\label{linearpredictionerror}
e_n^p\triangleq v_n-\widehat{v}_n^p
\end{equation}
is the estimation error of the linear estimator $\widehat{v}_n^p$ as in \eqref{linearestimation}. It should be emphasized that, in practice, we never have access to the true error $e_n^p$, but only to ${\widehat{e}}_n^{\,p}$, defined in \eqref{estimatedestimationerrors}. However, for simplicity of the exposition\footnote{The initial resolution parameter $\alpha_0$ can be chosen such that $\Pr\left({\widehat{e}}_n^{\,p}=e_n^p\right)$ is arbitrarily close to $1$. It can even be exactly equal to $1$ in case some (possibly partial) knowledge about the support of $x_n$ is available. At any rate, we touch upon and handle this aspect more accurately in the next subsection.}, and as mentioned above, we assume that $\widetilde{y}_n=v_n$, which means that ${\widehat{e}}_n^{\,p}=e_n^p$, during the entire initialization phase, hence $e_n^p$ appear in \eqref{LMSequation}.

In addition, rather than using $\{v_n\}$ \eqref{defofv}, we use the ``$\alpha_{n}$-standardized" process $\{\mybar{v}_n\}$ as the observations in \eqref{linearestimation}, since as the adaptive process evolves and $\alpha_n$ increases, the SOSs of $\mybar{v}_n$ gradually become less affected by $\alpha_n$ \eqref{powerofvtilde}. This alleviates the estimation (/learning) of the optimal filter coefficients. For a more detailed explanation, see Appendix \ref{AppA_vbar_vs_v}.

A discussion on the convergence of the LMS algorithm, as well as the how to choose the appropriate step size $\mu$ which guarantees this convergence, will be given in Subsection \ref{subsec:systemparameters}. For now, assume that $\mu$ is chosen so as to ensure that \cite{feuer1985convergence},
\begin{equation}\label{meanconvergencetoopt}
\alpha_n=\alpha_0:\;\lim_{n\to\infty} \Eset\left[\widehat{\uh}_n\right]=\uh_{\text{opt}},
\end{equation}
where $\uh_{\text{opt}}$ is the optimal length-$p$ filter corresponding to the oracle LMMSE estimator, a function of $\alpha_n=\alpha_0$ and $R_x[\ell]$.

\begin{algorithm}[t]
	\KwIn{$y_n, \widehat{\ubv}_n, \widehat{\uh}_n, R$}
	\KwOut{$\widehat{v}_n, \widehat{v}_n^p$}
	\nl Compute the linear estimate of $v_n$ based on $\ubv_n$
	\begin{equation}\label{linearestimation}
	\widehat{v}_n^p\triangleq\widehat{\uh}_n^{\tps}\widehat{\ubv}_n-\frac{1}{2},
	\end{equation}
	where the shift is to compensate for $\Eset[v_n]=\frac{1}{2}$ \eqref{defofv};\\
	\nl Compute
	\begin{align}
	w_n &= [y_n-\widehat{v}_n^p]\;{\rm{mod}}\;2^R,\\
	{\widehat{e}}_n^{\,p} &= \left(\left[w_n+\frac{1}{2}2^R\right]\;{\rm{mod}}\;2^R\right)-\frac{1}{2}2^R;\label{estimatedestimationerrors}
	\end{align}\\
	\nl return $\widehat{v}_n\triangleq\widehat{v}_n^p+\widehat{e}_n^{\,p},\;\widehat{v}_n^p$.
	\caption{{\bf Blind Modulo Unfolding} \newline $\widehat{v}_n,\,\widehat{v}_n^p=\text{BlindModUnfold}(y_n, \widehat{\ubv}_n, \widehat{\uh}_n, R)$ \label{Algorithm2}}
\end{algorithm}
\setlength{\textfloatsep}{2pt}
As an intermediate summary for the initialization, we have:
\tcbset{colframe=gray!90!blue,size=small,width=0.49\textwidth,halign=flush center,arc=2mm,outer arc=1mm}
\begin{tcolorbox}[upperbox=visible,colback=white,halign=left]
	\textbf{\underline{Phase 1: Initialization}. }\textbf{Input:} $\widehat{\uh}_0, \alpha_0$.
	\begin{enumerate}[1.]
		\setlength\itemsep{0.2em}
		\item Fix $\alpha_n=\alpha_0$, and accumulate $p+1$ samples $\{y_i\}_{i=1}^{p+1}$;
		\item Compute $\{\widetilde{y}_i=v_i\}_{i=1}^{p}$ as in \eqref{degeneratemodulo2}, and $\ubv_{p+1}$ as in \eqref{normalizedv};
		\item Set $\widehat{\uh}_{p+1}=\widehat{\uh}_0$;
		\item For $n=p+1,p+2,\ldots$ do
			\begin{enumerate}
			\setlength\itemsep{0.2em}
			\item[4.1.] $\widehat{v}_n,\,\widehat{v}_n^p=\text{BlindModUnfold}(y_n, \ubv_n, \widehat{\uh}_n, R)$;
			\item[4.2.] $\widehat{\uh}_{n+1}=\widehat{\uh}_{n}+\mu\cdot\ubv_ne_n^p$.
		\end{enumerate}
	\end{enumerate}
\end{tcolorbox}
After enough iterations, since we assume that $\alpha_0$ is sufficiently small to ensure that $\widehat{v}_n=v_n$ for every $n$ during initialization, which gives us access to $e_n^p$ \eqref{linearpredictionerror}, the filter $\widehat{\uh}_n$ will approximately converge to an unbiased estimate of $\uh_{\text{opt}}$, as in \eqref{meanconvergencetoopt}. Assuming the learning rate $\mu$ is sufficiently small, the MSE of $\widehat{v}^p_n$ will approximately converge to the MSE of $\widehat{v}_{\text{\tiny LMMSE},n}^p$ (with $\alpha_n$ replacing $\alpha$, according to the definition \eqref{defofv}),
\begin{equation}\label{defMSEoflinearest}
\exists N_0:\forall n>N_0:\Eset\left[\left(e_n^p\right)^2\right]\underset{\mu\ll1}{\approx}\sigma_{\text{\tiny LMMSE},p}^2.
\end{equation}
Accordingly, assuming that $v_n$ is sufficiently temporally predictable (due to $R_x[\ell]$), once $\Eset\left[\left(e_n^p\right)^2\right]$ is close enough to $\sigma_{\text{\tiny LMMSE},p}^2$, by virtue of \eqref{upperbound1}, an overload will not occur w.h.p., namely,
\begin{equation}\label{defOLofaaptive}
\Pr\left(\mathcal{E}_{\tiny {\rm{OL}}_n}\right)\triangleq\Pr\left(\left|e_{n}^p\right|\geq\frac{1}{2}2^R\right)\approx\Pr\left(\mathcal{E}^*_{\tiny {\rm{OL}}_n}\right),
\end{equation}
where $\mathcal{E}^*_{\tiny {\rm{OL}}_n}$ is defined in \eqref{nooverloadeventdefinition}, and refers to the overload event of the informed mod-ADC. In other words, the no overload event $\mathcal{E}_{\tiny \mybar{{\rm{OL}}}_n}\triangleq\left\{\widehat{e}_n^{\,p}=e_n^p\right\}$, which is the complement of the overload event $\mathcal{E}_{\tiny {\rm{OL}}_n}$, occurs w.h.p. At this point, we are ready to increase the resolution parameter $\alpha_n$, so as to decrease the effective modulo range $M_n$, and use the quantizer's output raw bits to a finer description of the input signal. This transition phase is described next.

\subsection{Phase 2: Updating the Resolution Parameter}\label{phase2adaptation}

As explained above, in order to increase $\alpha_n$, we must somehow detect that the filter $\widehat{\uh}_n$ already approximates the optimal one well enough, such that the induced estimation errors $e_n^p$ are sufficiently small with respect to the dynamic range $\Delta=2^R$. When this is the case, a small change in the resolution would not affect our ability to recover $v_n$ w.h.p.\ from $y_n$ and $\widehat{v}_n^p$. 

To see this, assume that we increase the resolution parameter $\alpha_{n}=\alpha_0+\epsilon_{\alpha}$, where $\epsilon_{\alpha}\in\Rset^+$ is a small increment, and accordingly also scale the respective filter coefficients by $\frac{\alpha_0+\epsilon_0}{\alpha_0}$. Now, $\widehat{\uh}_n$ is no longer optimal, since the second-order statistical properties of $\{\mybar{v}_{\ell}\}_{\ell\geq n}$ are different than those of $\{\mybar{v}_{\ell}\}_{\ell< n}$, based on which $\widehat{\uh}_n$ has been estimated thus far. However, if $\epsilon_{\alpha}$ is sufficiently small, a straightforward ``small-error" analysis yields that $\widehat{\uh}_n$ will now only slightly deviate from the (approximately) optimal filter to the new value of $\alpha_n$, such that right after increasing the resolution,
\begin{equation}\label{smallperturbations}
\alpha_{n}=\alpha_0+\epsilon_{\alpha} \; \Longrightarrow \; \widehat{\uh}_{n}=\widehat{\uh}^{\text{o}}_n+\uepsilon_n,
\end{equation}
where $\widehat{\uh}^{\text{o}}_n$ denotes an unbiased estimator of the optimal FIR filter corresponding to the LMMSE estimator of $v_n$ for the \emph{updated} resolution parameter $\alpha_0+\epsilon_{\alpha}$, and $\uepsilon_n\in\Rset^{p\times1}$ is a vector of ``small" biases, due to $\epsilon_{\alpha}$. Accordingly, the MSE of the (currently no longer optimal) linear estimator $\widehat{v}_n^p$, conditioned on the $p$ no overload past events\footnote{Strictly speaking, $\sigma_{\text{\tiny LMMSE},p}^2$ in \eqref{MSEoflinearestperturbation} of the blind mod-ADC is not equal to the MSE of \eqref{LMMSElengthp}, since for informed mod-ADC we do not condition on $\nooverloadnp$. However, under mild conditions, stated explicitly below, the difference is negligible, hence we use the same notation for simplicity.} $\nooverloadnp\triangleq\bigcap_{k=1}^p\mathcal{E}_{\tiny \mybar{{\rm{OL}}}_{n-k}}$ enabling the exact recovery of $v_n$, will be slightly increased\footnote{For simplicity, we assume here that prior to increasing the resolution $\sigma_{p,n}^2=\sigma_{\text{\tiny LMMSE},p}^2$, such that $\sigma_{\varepsilon}^2$ is \emph{only} due to $\epsilon_{\alpha}$. In practice, we have $\sigma_{p,n}^2\approx\sigma_{\text{\tiny LMMSE},p}^2$, such that $\sigma_{\varepsilon}^2$ encapsulates estimation errors due to $\widehat{\uh}_n$ as well. Still, after changing $\alpha_n$, $\sigma_{\varepsilon}^2$ will be dominated by errors due to $\epsilon_{\alpha}.$}
\begin{equation}\label{MSEoflinearestperturbation}
\sigma^2_{p,n}\triangleq\Eset\left[\left.\left(e_n^p\right)^2\right|\nooverloadnp\right]=\sigma_{\text{\tiny LMMSE},p}^2+\underbrace{\sigma_{\varepsilon}^2}_{\text{Due to $\epsilon_{\alpha}$}}.
\end{equation}
Nonetheless, as long as $\sigma_{p,n}\ll\frac{1}{2}2^R$, such that the event $\mathcal{E}_{\tiny \mybar{{\rm{OL}}}_n}$ still occurs w.h.p., $v_n$ is still exactly recovered w.h.p.\ using Algorithm \ref{Algorithm2}. Indeed, an important observation is that $v_n$ can be recovered using Algorithm \ref{Algorithm2} even when a suboptimal linear  estimator is used in \eqref{linearestimation}. For successful operation, we \emph{only} require that the linear estimator would be accurate enough to ensure that $\mathcal{E}_{\tiny \mybar{{\rm{OL}}}_n}$ occurs w.h.p.
Consequently, for short transition periods in which the optimal filter is learned, a suboptimal filter would suffice.

Hence, we conclude the following:
\begin{enumerate}[i.]
\item[\circled{1}] If the resolution parameter is adapted in small increments, we are able to maintain sufficiently small estimation errors, and safely continue recovering $v_n$ w.h.p.; and
\item[\circled{2}] Before increasing the resolution, we desire to arrive at an intermediate steady state, wherein $\mathcal{E}_{\tiny \mybar{{\rm{OL}}}_n}$ occurs w.h.p.
\end{enumerate}
Since $\epsilon_{\alpha}$ is a user-controlled system parameter, $\circled{$1$}$ can be easily achieved. As for $\circled{$2$}$, since we are operating in a blind scenario, where the distribution of the input $x_n$ is unknown, it is generally unclear how to ensure rarity of no overload. Therefore, for this purpose only, we take the simplifying, but useful, assumption that $\left.e_n^p\right|\nooverloadnp\sim\mathcal{N}(0,\sigma_{p,n}^2)$. Note, however, that this assumption is not strictly required in order to analytically justify the derivation which follows, and is merely to simplify the exposition. In this case, $\sigma_{p,n}$ is directly related to $\mathcal{E}_{\tiny \mybar{{\rm{OL}}}_n}$, conditioned on $\nooverloadnp$. Specifically, if $\Delta/2=\kappa\cdot\sigma_{p,n}$ for some $\kappa\in\Rset^+$, then we have
\begin{equation}\label{boundgaussianQfunc}
\Pr\left(\left.|e_n^p|>\frac{1}{2}2^R\right|\nooverloadnp\right)=2Q\left(\frac{\Delta}{2\sigma_{p,n}}\right)=2Q\left(\kappa\right),
\end{equation}
where $Q(x)\triangleq\int_{x}^{\infty}\frac{1}{\sqrt{2\pi}}e^{-t^2/2}{\rm{d}}t$ is the $Q$-function. Put simply, if the linear estimator is good enough, such that half the modulo range $\Delta/2$ is $\kappa$ times greater than its Root MSE (RMSE), and $\kappa$ is sufficiently large, $\mathcal{E}_{\tiny \mybar{{\rm{OL}}}_n}|\nooverloadnp$ occurs w.h.p.\ This provides the conditions to re-learn the optimal filter corresponding to the LMMSE estimator of $v_n$ with the updated resolution $\alpha_{n}$.

In practice, though, since $R_x[\ell]$ is unknown, $\sigma^2_{p,n}$ is clearly not known as well. Nevertheless, since $\widehat{v}_n\overset{\text{w.h.p.}}{=}v_n$ throughout the adaptive process, we can estimate $\sigma_{p,n}$ online by
\begin{align}\label{eststdoflinearest}
{\widehat{\sigma}}^2_{p,n}\triangleq\frac{1}{L_s}\sum_{k=0}^{L_s-1}\left({\widehat{v}}_{n-k}-{\widehat{v}}^p_{n-k}\right)^2,\; {\widehat{\sigma}}_{p,n}\triangleq\sqrt{{\widehat{\sigma}}^2_{p,n}}.
\end{align}
where $L_s\in\Nset^+$ is a moving average window length, and is also set to be the minimal (discrete) time stabilization interval wherein $\alpha_n$ must be kept fixed after its last update. More details on the system parameters $L_s$ and $\kappa$ are given in Subsection \ref{subsec:systemparameters}.
Thus, to achieve $\circled{$2$}$, we increase $\alpha_{n}$ only when $\mathbbm{1}_{\uparrow\alpha,n}=1$, where
\begin{equation}\label{indicatorkappastd}
\mathbbm{1}_{\uparrow\alpha,n} \triangleq \begin{cases}
1, & \kappa\cdot\widehat{\sigma}_{p,n}<\frac{\Delta}{2}\\
0, & \kappa\cdot\widehat{\sigma}_{p,n}>\frac{\Delta}{2}
\end{cases}.
\end{equation}
Whenever $\mathbbm{1}_{\uparrow\alpha,n}=0$, we infer that the estimation errors are not satisfactorily small. In these cases, we decrease the resolution so as to resort to a state where $v_n$ is again recovered w.h.p. By this, we allow the LMS filter \eqref{LMSequation} to converge to the desired filter, and then safely increase the resolution again. Following our previous observation on the accuracy required by the linear estimator $\widehat{v}_n^p$, at this point the desired filter is not necessarily the optimal one, but is merely one attaining $\mathbbm{1}_{\uparrow\alpha,n}=1$. Accordingly, it is certainly possible that $\alpha_n$ would be increased before $\widehat{\uh}_n$ would converge to the optimal filter (at least before steady state, as discussed in the next subsections).

Note that, conditioned on $\mathcal{E}_{\tiny \mybar{{\rm{OL}}}_{n+1}}^{(L_s)}$, \eqref{eststdoflinearest} is a consistent estimate (with respect to $L_s$) of $\sigma_{p,n}$ for a wide class of signals, even when the errors $\left.e_n^p\right|\nooverloadnp$ are non-Gaussian. Hence, this mechanism is generally robust, and relies on Gaussianity only for \eqref{boundgaussianQfunc}. Naturally, this assumption implies that the expected stability would be obtained for sub-Gaussian\footnote{The real-valued random variable $u$ is called \emph{sub-Gaussian} if $\exists c,\gamma\in\Rset^+$ such that $\forall t>0: \Pr(|u|>t)\leq c\cdot\exp\left\{-\gamma t^2\right\}$.} errors as well.

Upon updating $\alpha_n$, we also appropriately update the filter $\widehat{\uh}_n$, since the input signal $\ubv_n$ is scaled with $\alpha_n$ as well \eqref{normalizedv}. Therefore, it is convenient to use \emph{multiplicative} updates, rather than additive, to $\alpha_n$ and $\widehat{\uh}_n$. For a fixed $\delta_{\alpha}\in(0,1)$, we update
\begin{align}\label{updatealphaandfilter}
{\rm{Increase\;Resolution:}}\;\alpha_n&=\frac{\alpha_{n-1}}{\delta_{\alpha}},\; \widehat{\uh}_n=\frac{1}{\delta_{\alpha}}\widehat{\uh}_{n-1},\\
{\rm{Decrease\;Resolution:}}\;\alpha_n&=\delta_{\alpha}\alpha_{n-1},\; \widehat{\uh}_n=\delta_{\alpha}\widehat{\uh}_{n-1}.
\end{align}
A summary for the resolution updating phase is as follows:
\tcbset{colframe=gray!90!blue,size=small,width=0.49\textwidth,halign=flush center,arc=2mm,outer arc=1mm}
\begin{tcolorbox}[upperbox=visible,colback=white,halign=left]
	\textbf{\underline{Phase 2: Resolution Update}. }\textbf{Input:} $\kappa, \delta_{\alpha}$.\\
	Assumption: $\alpha_n$ was held fixed (at least) $L_s$ samples.
	\begin{enumerate}[1.]
		\setlength\itemsep{0.2em}
		\item $\alpha_{n+1}=\left[\mathbbm{1}_{\uparrow\alpha,n}\cdot\frac{1}{\delta_{\alpha}}+\left(\mathbbm{1}_{\uparrow\alpha,n}-1\right)\cdot\delta_{\alpha}\right]\alpha_{n}$;
		\item $\widehat{\uh}_{n+1}=\left[\mathbbm{1}_{\uparrow\alpha,n}\cdot\frac{1}{\delta_{\alpha}}+\left(\mathbbm{1}_{\uparrow\alpha,n}-1\right)\cdot\delta_{\alpha}\right]\widehat{\uh}_{n}$.
	\end{enumerate}
\end{tcolorbox}

It is now straightforward to generalize this adaptive process, since, conceptually, we now only need to repeatedly execute the properly interlaced Phase 1 and Phase 2. In the repeated Phase 1, the ``initial" values for the filter and resolution parameter would be the corresponding values of the previous time step. Additionally, $\ubv_n$ will be replaced by $\widehat{\ubv}_n\triangleq\left[\widehat{\mybar{v}}_{n-1}\, \cdots\, \widehat{\mybar{v}}_{n-p}\right]^{\tps}\in\Rset^{p\times1}$, whose entries
\begin{equation}\label{defofvbarest}
\widehat{\mybar{v}}_{n}\triangleq\widehat{x}_n=\frac{\widehat{v}_n+\frac{1}{2}}{\alpha_{n}},
\end{equation}
are $\{\widehat{\mybar{v}}_{n-i}\overset{\text{w.h.p.}}{=}\mybar{v}_{n-i}\}_{i=1}^p$, assuming previous successful recoveries w.h.p.. The repeated Phase 2 would then be executed after (at least) $L_s$ time steps with the updated resolution.

Note that we intentionally use in \eqref{defofvbarest}, and hereafter, the (seemingly redundant) notation $\widehat{\mybar{v}}_{n}$ rather than $\widehat{x}_n$, since in this context we are actually trying to perfectly recover $\mybar{v}_n$, defined in \eqref{normalizedv}, rather than to estimate $x_n$. This is to enable the proper operation of the LMS algorithm \eqref{LMSequation}, whose input should be $\{\mybar{v}_{n-i}\}_{i=1}^p$, and not $\{x_{n-i}\}_{i=1}^p$ \cite{zamir2008achieving,ordentlich2018modulo}.

Ideally, alternating between these two phases would lead to convergence near the limit $\kappa\cdot\sigma_{p,n}=\frac{\Delta}{2}$, as in \eqref{indicatorkappastd}, up to small fluctuations due to the limited-resolution adaptation step $\delta_{\alpha}$ and estimation errors in $\widehat{\sigma}_{p,n}$. However, recall that $\left.\mathcal{E}_{\tiny \mybar{{\rm{OL}}}_n}\right|\nooverloadnp$, which implies that $\widehat{e}_n^p=e_n^p$, and in turn $\widehat{v}_n=v_n$, is \emph{only} w.h.p., and in practice this is certainly not true for all $n\in\Nset^+$. Indeed, whenever $\mathcal{E}_{\tiny {\rm{OL}}_n}=\{\widehat{v}_n\neq v_n\}$ occurs, an extremely fast and destructive error propagation process begins. To detect such errors and prevent the consequent error propagation, we propose the defense mechanism presented next.

\subsection{Error Propagation Prevention}\label{subsec:errorpropagation}
One natural way of coping, and eventually preventing the aforementioned error propagation is by splitting the problem into two parts. The first part is to detect that an error has occurred, namely that $v_n$ has not been perfectly recovered. In other words, the event $\mathcal{E}_{\tiny {\rm{OL}}_n}=\{\widehat{v}_n\neq v_n\}$ has to be detected, and as quickly as possible. The second part is, given that $\mathcal{E}_{\tiny {\rm{OL}}_n}$ has been detected, to mitigate the error effect so as to reclaim a high-resolution functioning mod-ADC steady state.

Provided that an error event has been detected, a simple, though conservative mitigation solution is to fully ``re-open" the effective modulo range \eqref{effectivemodulorange} to its initial value $M_0$ for (at least) $p$ time steps. By this, we effectively initialize the process and guarantee that no errors occur, at the expense of (locally) retreating to a low-resolution regime. This solution, however, is useful only if the detection of $\mathcal{E}_{\tiny {\rm{OL}}_n}$ can be handled very accurately, i.e., with a very low false-alarm rate. Otherwise, the average operational time percentage of the mod-ADC in a degenerate modulo state (corresponding to a large $M_n$) would be high, and there would be no gain in using a mod-ADC. Hence, we turn our attention to the detection of $\mathcal{E}_{\tiny {\rm{OL}}_n}$.

Formally, our goal now is to derive an estimator $\widehat{\mathbbm{1}}_{\mathcal{E}_{\tiny {\rm{OL}}_n}}\in\{0,1\}$ for the oracle indicator
\begin{equation}\label{oracledetector}
\mathbbm{1}_{\mathcal{E}_{\tiny {\rm{OL}}_n}} \triangleq \begin{cases}
1, & \mathcal{E}_{\tiny {\rm{OL}}_n}\\
0, & \mathcal{E}_{\tiny \mybar{{\rm{OL}}}_n}
\end{cases} = \begin{cases}
1, & {\widehat{v}}_{n}\neq v_n\\
0, & {\widehat{v}}_{n}= v_n
\end{cases}, \; \forall n\in\Nset^+.
\end{equation}
Since this is required at every time $n$, and assuming that with $M_n=M_0$ there are no overload events, this is essentially a change detection problem (e.g., \cite{aminikhanghahi2017survey}). In particular, since $\left.\mathcal{E}_{\tiny {\rm{OL}}_n}\right|\nooverloadnp$ ($\approx\mathcal{E}^*_{\tiny {\rm{OL}}_n}$, \eqref{defOLofaaptive}) is rare \eqref{upperbound1}, this problem can be viewed as a special instance of the fraud detection problem \cite{fawcett1997adaptive}, where $\widehat{v}_n$ is pretending to be $v_n$, while in fact it is not, viz., $\widehat{v}_n\neq v_n$.

Fortunately, our specific problem has favorable properties that allow us to develop an accurate detector. In particular, observe that increasing $\alpha_n$ essentially ``pushes" $\mybar{v}_n$ towards approximate wide-sense stationarity. Specifically, using \eqref{normalizedv},
\begin{equation}\label{approximatewss}
\Eset\left[\mybar{v}_n\mybar{v}_{n-\ell}\right]=R_x[\ell]+\frac{1}{12\alpha_{n}^2}\cdot\mathbbm{1}_{\ell=0},
\end{equation}
such that even if $\alpha_n$ changes over time, for sufficiently large values of $\alpha_n$, the autocorrelation of $\mybar{v}_n$---even when unknown---can be considered as being approximately a function of $\ell$ only. Furthermore, it is seen from \eqref{approximatewss} that the variance of $\mybar{v}_n$ \eqref{powerofvtilde} is the only source of non-stationarity.

Similarly to our comment above \eqref{boundgaussianQfunc}, in a blind scenario like the one under consideration here, information such as \eqref{approximatewss} is not necessarily sufficient in order to be able to design an accurate detector of the event $\mathcal{E}_{\tiny {\rm{OL}}_n}$. Therefore, at this point we again invoke Gaussianity, and assume that $\{x_n\}$ is a Gaussian process with an autocorrelation function $R_x[\ell]$.

For this case, it is known that \cite{leadbetter2012extremes}, if
\begin{equation}\label{condonautocorr}
R_x[\ell]\cdot\log(\ell)\xrightarrow{\ell\to\infty}0,
\end{equation}
then
\begin{equation}\label{maxvalupperbound}
\Pr\left(\|x_n\|_{\ell_N^{\infty}}\leq\sqrt{2\sigma_x^2\log(N)}\right) \xrightarrow{N\to\infty}1,
\end{equation}
where $\|x_n\|_{\ell_N^{\infty}}\triangleq\sup_{n\in\{1,\ldots,N\}}|x_n|$. Since $\{z_n\in(-1,0]\}$ is a process with bounded support, assuming that $\min_n\{\alpha_n\}=\alpha_0$, we also have
\begin{equation}\label{maxvalupperboundonvbar}
\Pr\left(\|\mybar{v}_n\|_{\ell_N^{\infty}}\leq\sqrt{2\left(\sigma_x^2+\frac{1}{12\alpha_0^2}\right)\log(N)}\right) \xrightarrow{N\to\infty}1.
\end{equation}

Recall, however, that $\{\alpha_n\}$ is typically an increasing sequence on average, and conditioned on no overload events $\mathcal{E}_{\tiny \mybar{{\rm{OL}}}_{n+1}}^{(n)}$, converges (up to small fluctuations) to the value for which $\kappa\cdot\sigma_{p,n}=\frac{\Delta}{2}$, as explained in the previous subsection. Thus, under $\bigcap_{k=1}^n\{\mathbbm{1}_{\mathcal{E}_{\tiny {\rm{OL}}_{k}}}=0\}$, in the absence of estimation errors in $\widehat{\sigma}_{p,n}$ and with an infinite resolution step size  $\delta_{\alpha}\to1$,
\begin{equation}\label{limitalphadef}
\lim_{n\to\infty}\alpha_n\triangleq\alpha_{\infty}\;\Rightarrow\;\lim_{n\to\infty}\Eset\left[\mybar{v}_n^2\right]=\sigma_x^2+\frac{1}{12\alpha_{\infty}^2}\triangleq\sigma_{\bar{v}}^2.
\end{equation}
Hence, for the ideal steady state process $\mybar{v}_n^{\infty}\triangleq(v_n+\frac{1}{2})/\alpha_{\infty}$, we have
\begin{equation}\label{maxvalupperboundonvbarinf}
\Pr\left(\|\mybar{v}_n^{\infty}\|_{\ell_N^{\infty}}\leq\sqrt{2\sigma_{\bar{v}}^2\log(N)}\right) \xrightarrow{N\to\infty}1,
\end{equation}
or, equivalently,
\begin{equation}\label{maxvalupperbound2ndform}
\Pr\left(\|\mybar{v}_n^{\infty}\|_{\ell_N^{\infty}}>\sqrt{2\sigma_{\bar{v}}^2\log(N)}\right) \xrightarrow{N\to\infty}0.
\end{equation}
This means that, asymptotically, knowing only the variance of the process $\mybar{v}_n$, and observing its magnitudes, is \emph{sufficient} in order to detect almost surely an abnormality in the form of a large, improbable deviation exceeding the threshold in \eqref{maxvalupperbound2ndform}.

Now, observe that an overload event $\mathcal{E}_{\tiny {{\rm{OL}}}_n}$ inflicts an estimation error in $\widehat{v}_n$, and in turn in $\widehat{\mybar{v}}_n$, of the order of $\Delta$. Clearly, this creates a large ``discontinuity"\footnote{This, of course, is not a discontinuity in the formal sense as defined for deterministic functions. Rather, we use this term here informally to refer to an improbable transition from one value to another, in a manner that is inconsistent with $R_x[\ell]$ and \eqref{maxvalupperboundonvbarinf}, governing the statistical nature of $\{\mybar{v}_n\}$}, which is exactly the abnormality form we can identify w.h.p.\ according to \eqref{maxvalupperbound2ndform}. In light of all the above, we propose
\begin{equation}\label{ourdetector}
\widehat{\mathbbm{1}}_{\mathcal{E}_{\tiny {\rm{OL}}_n}} \triangleq \begin{cases}
1, & |\widehat{\mybar{v}}_n|>\sqrt{2\widehat{\sigma}_{\bar{v},n}^2\log(n)}\\
0, & \text{otherwise}
\end{cases}, \; \forall n\geq N_s,
\end{equation}
as the detector of an error event due to $\mathcal{E}_{\tiny {{\rm{OL}}}_n}$, where 
\begin{equation}\label{varianceestofvbar}
\widehat{\sigma}_{\bar{v},n}^2\triangleq\frac{1}{n-1}\sum_{k=1}^{n-1}\widehat{\mybar{v}}_{k}^2,\;\; \forall n\geq 2
\end{equation}
and $N_s\in\Nset^+$ is a fixed stabilization time-interval, wherein \eqref{ourdetector} is still not sufficiently accurate, and we enforce a simple, more conservative condition for the transition phase $n\leq N_s$. For example, one reasonable choice could be
\begin{equation}\label{ourdetectorinitial}
\widehat{\mathbbm{1}}_{\mathcal{E}_{\tiny {\rm{OL}}_n}} \triangleq \begin{cases}
1, & |\widehat{\mybar{v}}_n|>\beta\cdot\widehat{\sigma}_{\bar{v},n}\\
0, & \text{otherwise}
\end{cases}, \; \forall n< N_s,
\end{equation}
where $\beta\in\Rset^+$ is some predefined number (e.g., $\beta=5$). From practical considerations, since the threshold value in \eqref{ourdetector} increase logarithmically with $n$, a plausible solution would be to reset the time-index in this threshold every error event $\mathcal{E}_{\tiny {\rm{OL}}_n}$.

Once we observe $\widehat{\mathbbm{1}}_{\mathcal{E}_{\tiny {\rm{OL}}_n}}=1$, we set $\alpha_{n+1}=\alpha_0$, and reset the process as described above, in the beginning of this subsection.
The proposed error propagation defense mechanism is summarized as follows:
\tcbset{colframe=gray!90!blue,size=small,width=0.49\textwidth,halign=flush center,arc=2mm,outer arc=1mm}
\begin{tcolorbox}[upperbox=visible,colback=white,halign=left]
	\textbf{\underline{Error Propagation Defense Mechanism} }\\
	Assumption: $n\geq p$.
	\begin{enumerate}[1.]
		\setlength\itemsep{0.2em}
		\item Update $\widehat{\sigma}_{\bar{v},n}^2=\frac{n-2}{n-1}\cdot\widehat{\sigma}_{\bar{v},n-1}^2+\frac{1}{n-1}\cdot\widehat{\mybar{v}}_{n-1}^2$;
		\item If $\widehat{\mathbbm{1}}_{\mathcal{E}_{\tiny {\rm{OL}}_n}}=1$
		\begin{enumerate}
			\setlength\itemsep{0.2em}
			\item[2.1.] Reset $\alpha_{n+1}=\alpha_0$, and adapt $\widehat{\uh}_{n+1}=\frac{\alpha_0}{\alpha_{n}}\cdot\widehat{\uh}_{n}$.
		\end{enumerate}
	\end{enumerate}
\end{tcolorbox}
We note that, conditioned on the no overload event $\mathcal{E}_{\tiny \mybar{{\rm{OL}}}_n}$, and assuming \eqref{limitalphadef} holds, \eqref{varianceestofvbar} is consistent, namely,
\begin{equation*}
\widehat{\sigma}_{\bar{v},n}^2\xrightarrow[\quad\;]{P}\sigma_{\bar{v}}^2,
\end{equation*}
where $\xrightarrow[\quad\;]{P}$ denotes convergence in probability as $n\to\infty$. Thus, the decision rule in \eqref{ourdetector} becomes increasingly accurate as we approach steady state, indeed, a desirable outcome.

\subsection{Key System Parameters and Corresponding Trade-offs}\label{subsec:systemparameters}
First and foremost, convergence of the adaptive process described above is conditioned on the no overload event. Therefore, the parameter $\kappa$, dictating the desired confidence level in which the estimation errors $e_n^p$ are kept inside $\left(-\frac{\Delta}{2},\frac{\Delta}{2}\right)$, must be set to a sufficiently large value, so as to ensure that \eqref{boundgaussianQfunc} is sufficiently low. For example, choosing $\kappa=7$ already gives $\Pr\left(\left.\mathcal{E}_{\tiny {\rm{OL}}_n}\right|\nooverloadnp\right)\approx2.5596\times10^{-12}$. Yet, as $\kappa$ increases, the asymptotic resolution $\alpha_{\infty}$ of the blind mod-ADC decreases, as already alluded from \eqref{indicatorkappastd}. A formal characterization of this asymptotic trade-off is provided in the next subsection.

Given that $\kappa$ was chosen properly, we continue with the convergence and asymptotic analysis, conditioned on no overload. Specifically, we now focus on the learning rate $\mu$. Assuming momentarily that $\alpha_n$ is fixed, based on the well-established theory of the LMS algorithm \cite{benveniste2012adaptive}, if we choose
\begin{equation}\label{mucondition}
0<\mu<\frac{1}{\Tr\left(\Eset\left[\mybar{\uv}_n\mybar{\uv}^{\tps}_n\right]\right)}=\frac{1}{p\cdot(\sigma_x^2+\frac{1}{12\alpha_{n}^2})},
\end{equation}
then the FIR filter $\widehat{\uh}_n$ would converge in the sense \eqref{meanconvergencetoopt}, namely it will randomly fluctuate about $\uh_{\text{opt}}$, corresponding to the LMMSE estimator. Recall that conditioned on no overload, $\widehat{\mybar{\uv}}_n=\mybar{\uv}_n$, and when $\alpha_{n}$ is fixed, $\mybar{\uv}_n$ is stationary in the respective time interval, hence the diagonal elements of $\Eset\left[\mybar{\uv}_n\mybar{\uv}^{\tps}_n\right]\in\Rset^{p\times p}$ are all equal to the variance \eqref{powerofvtilde}, and the right hand side of \eqref{mucondition} follows. Now, since $\alpha_{n}$ is in fact time-varying, and is typically an increasing sequence on average, we would like to choose $\mu$ such that
\begin{equation}\label{mucondition2}
0<\mu<\frac{1}{p\cdot(\sigma_x^2+\frac{1}{12\min_n\{\alpha_{n}^2\}})}=\frac{1}{p\cdot(\sigma_x^2+\frac{1}{12\alpha_{0}^2})}.
\end{equation}

However, the upper bound \eqref{mucondition2} is unknown, since $\sigma_x^2$ is unknown. Therefore, we propose to choose
\begin{equation}\label{muourproposition}
\mu=\frac{\epsilon_{\mu}}{p\cdot\widehat{\sigma}_{\bar{v},p+1}^2},
\end{equation}
where $\epsilon_{\mu}\in\Rset^+$ is some small constant (e.g., $10^{-2}$), and $\widehat{\sigma}_{\bar{v},p+1}^2$ \eqref{varianceestofvbar} can be computed during the initialization phase. Since $\alpha_0^{-2}$ is typically small, and thus $\sigma_{\bar{v},p+1}^2$ is dominated by $\sigma_x^2$, our empirical experience indicates that choosing $\epsilon_{\mu}$ appropriately, so as to ensure the desired stability, is rather easy. Furthermore, a longer initialization phase (i.e., more than $p$ discrete-time steps) could be performed, which would yield a more accurate estimate of the variance of $\mybar{v}_n$. Lastly, and although not necessary, $\mu$ could be easily adapted throughout the process based on the online estimate \eqref{varianceestofvbar}.

Another system parameter is $L_s$, the minimal discrete-time interval in which the resolution $\alpha_n$ must be held fixed before another resolution update is allowed. In the extreme case $L_s\to\infty$, we have the highest stability ($\alpha_n$ is fixed, $\widehat{\uh}_n$ converges) but the slowest (zero) progress towards high resolution. In the other extreme case $L_s=1$, $\alpha_n$ can be updated at all times, but the estimate \eqref{eststdoflinearest}, and therefore the detector \eqref{indicatorkappastd}, become extremely inaccurate. Therefore, $L_s$ should be set so as to appropriately handle this trade-off. Since Algorithm \ref{Algorithm2} assumes that the previous $p$ samples of $\mybar{v}_n$ are available (via $\widehat{\mybar{v}}_n$), it is reasonable to choose $L_s$ proportional to $p$ (e.g., ${\rm{round}}(p/2)$).

The resolution step size parameter $\delta_{\alpha}\in(0,1)$ should also balance a similar trade-off. As $\delta_{\alpha}$ decreases, the convergence towards $\alpha_{\infty}$ is faster. However, the LMS would be required to cope with more abrupt changes in the variance of $\ubv_n$, harming the linear estimator $\widehat{\uv}_n^p$, and thus locally inflicting larger estimation errors $e_n^p$, which could lead to an overload event. On the other hand, as $\delta_{\alpha}$ approaches $1$ (from below), the transition becomes smoother, allowing the LMS to adjust conveniently, and based on the same principles explained above, decrease the probability of an overload event. Of course, this comes at the cost of a slower convergence rate to $\alpha_{\infty}$.

To conclude this section, we refer to the parameter $p$, the length of the FIR filter producing the linear estimator $\widehat{v}_n^p$. Preferably, $p$ should be chosen based on some prior knowledge related to the specific application for which the mod-ADC is being used. In particular, if the effective support\footnote{For some $\epsilon>0$, the $\epsilon$-effective support of $R_x[\ell]$ is the number $L_x\in\Nset^+$ for which $\forall|\ell|>L_x:|R_x[\ell]|<\epsilon$. Loosely speaking, we say that $L_x$ is the effective support of $R_x[\ell]$ when $\forall|\ell|>L_x:|R_x[\ell]|\approx0$.} $L_x$ of the unknown autocorrelation function $R_x[\ell]$ is known even approximately, then an educated choice would be $p=L_x$. Indeed, if the support of $R_x[\ell]$ is \emph{precisely} $L_x$, then the causal Wiener filter \cite{kamen1999wiener}, i.e., the optimal (generally not FIR) filter corresponding to the LMMSE estimator, for estimating the process $\{v_n\}$ based on $\{\mybar{v}_n\}$ is an FIR filter of length $p=L_x+1$ (the ``$+1$" is due to the present sample, but \eqref{linearestimation} uses only past samples). We note in passing that $L_x$ can be estimated during the initialization phase, since $\widetilde{y}_n=\alpha_{n}x_n+z_n=v_n$ as long as $\alpha_n=\alpha_0$, and $\alpha_0$ can always be chosen so as to ensure the equality $\widetilde{y}_n=v_n$ (w.h.p.).

\subsection{The Asymptotic Performance of a Blind Mod-ADC}\label{subsec:convergenceasymptotictradeoff}
Let us assume that all the parameters have been chosen such that an overload does not occur. In this ideal (merely theoretical) case, if we assume further that $\widehat{\sigma}^2_{p,n}=\sigma^2_{p,n}$ and that an infinitely fine step size $\delta_{\alpha}\to1$ is used, the resolution of the blind mod-ADC converges to $\alpha_{\infty}$ \eqref{limitalphadef}. In this asymptotic state, we have the equilibrium
\begin{equation}\label{equilibrium}
\kappa\cdot\widetilde{\sigma}_{p,\infty}=\frac{\Delta}{2},
\end{equation}
where
\begin{equation}\label{definitionofasymptoticrmse}
\widetilde{\sigma}^2_{p,\infty}\triangleq\lim_{n\to\infty}\widetilde{\sigma}^2_{p,n}, \;\;\widetilde{\sigma}^2_{p,n}\triangleq\Eset\left[\left.\left(e^p_n\right)^2\right|\mathcal{E}_{\tiny \mybar{{\rm{OL}}}_{n+1}}^{(n)}\right],
\end{equation}
and notice the difference between $\widetilde{\sigma}^2_{p,n}$ in \eqref{definitionofasymptoticrmse} and $\sigma^2_{p,n}$ in \eqref{MSEoflinearestperturbation}. Now, recall that $\widehat{v}_n^p$ is a function of $\ubv_n$ \eqref{linearestimation}, which, asymptotically, is a function of $\alpha_{\infty}$. Hence $\widetilde{\sigma}^2_{p,\infty}$ is also a function of $\alpha_{\infty}$. Thus, we conclude that under the ideal (only theoretical) conditions mentioned above, the highest resolution attainable for a particular fixed set of system parameters (e.g., $\kappa, p, \Delta=2^R$) is governed by the equilibrium equation \eqref{equilibrium}.

Of course, in practice, both $\widehat{\sigma}^2_{p,n}\neq\sigma^2_{p,n}$ and the occurrence of an overload event at some point are with probability $1$, and at any rate $\delta_{\alpha}$ is obviously finite. Nevertheless, we now know that even under the best theoretical conditions, for a particular set of system parameters, the highest resolution is limited. This motivates us to identify the point in time at which the system has reached its limiting capability, and stop the resolution adaptation, favoring stability---which yields stationarity from that point onwards---and reducing the computational load. Clearly, the optimal scenario is the one in which the adaptation-free mod-ADC is working at all times at the highest attainable resolution $\alpha_{\infty}$.

Fortunately, it is actually possible to detect the equilibrium \eqref{equilibrium} quite accurately as follows. Define the linear estimator of the ``$\alpha_{n}$-standardized" process $\mybar{v}_n$,
\begin{equation}\label{vbarlinearest}
\widehat{\mybar{v}}_n^p\triangleq\frac{\widehat{v}_n^p+\frac{1}{2}}{\alpha_n},\quad \forall n\geq p+1.
\end{equation}
This estimator has the following conditional MSE,
\begin{align}\label{MSEvbarlinearest}
\mybar{\sigma}_{p,n}^2&\triangleq\Eset\left[\left.\left(\mybar{v}_n-\widehat{\mybar{v}}_n^p\right)^2\right|\mathcal{E}_{\tiny \mybar{{\rm{OL}}}_{n+1}}^{(n)}\right]\\
&=\frac{1}{\alpha_{n}^2}\Eset\left[\left.\left(v_n-\widehat{v}_n^p\right)^2\right|\mathcal{E}_{\tiny \mybar{{\rm{OL}}}_{n+1}}^{(n)}\right]=\frac{\widetilde{\sigma}^2_{p,n}}{\alpha_n^2}.
\end{align}
Therefore, assuming the same ideal theoretical conditions as described above hold, asymptotically,
\begin{gather}\label{asymptoticMSEvbarlinearest}
\mybar{\sigma}_{p,\infty}^2\triangleq\lim_{n\to\infty}\mybar{\sigma}_{p,n}^2=\lim_{n\to\infty}\frac{\widetilde{\sigma}^2_{p,n}}{\alpha_n^2}=\frac{\widetilde{\sigma}^2_{p,\infty}}{\alpha^2_{\infty}}\\
\Longrightarrow\;\widetilde{\sigma}_{p,\infty}=\alpha_{\infty}\cdot\mybar{\sigma}_{p,\infty}.\label{LMMSEasalpha}
\end{gather}
Substituting \eqref{LMMSEasalpha} into \eqref{equilibrium} gives the equivalent equilibrium
\begin{equation}\label{equilibriumwithalpha}
\mybar{\sigma}_{p,\infty}=\frac{1}{2\kappa}\cdot\frac{\Delta}{\alpha_{\infty}}\triangleq\frac{M_{\infty}}{2\kappa}.
\end{equation}

In view of \eqref{equilibriumwithalpha}, we propose the following
\tcbset{colframe=gray!90!blue,size=small,width=0.49\textwidth,halign=flush center,arc=2mm,outer arc=1mm}
\begin{tcolorbox}[upperbox=visible,colback=white,halign=left]
	\textbf{\underline{Steady State Detector} }\\
	\begin{equation}\label{steadystatedetector}
	\widehat{\mathbbm{1}}_{M_{\infty},n} \triangleq \begin{cases}
	1, & {\widehat{\mybar{\sigma}}}_{p,n}>\frac{M_n}{2\kappa}\\
	0, & \text{otherwise}
	\end{cases}, \; \forall n\geq N_s,
	\end{equation}
\end{tcolorbox}
\noindent where 
\begin{equation}\label{sigmabarest}
\widehat{\mybar{\sigma}}_{p,n}^2\triangleq\frac{1}{n}\sum_{k=1}^{n}\left(\widehat{\mybar{v}}_{k}-\widehat{\mybar{v}}_{k}^p\right)^2,\;\; \forall n\in\Nset^+.
\end{equation}
In words, when the estimated RMSE of the linear estimator $\widehat{\mybar{v}}_n^p$ is $\kappa$ times greater than half the effective modulo range, we estimate that the mod-ADC has reached the limit of its capability, in terms of the highest attainable resolution for the given set of system parameters. Note the difference between \eqref{indicatorkappastd} and \eqref{steadystatedetector}, where the former uses $\widehat{\sigma}_{p,n}$ and the latter uses $\widehat{\mybar{\sigma}}_{p,n}$, respectively. As seen from its definition \eqref{eststdoflinearest}, $\widehat{\sigma}_{p,n}$ is a ``short-term" memory estimate of the ``local" standard deviation of $v_n$. In contrast, as seen from \eqref{sigmabarest}, $\widehat{\mybar{\sigma}}_{p,n}$ is a ``long-term" memory estimate of the average standard deviation of $\mybar{v}_n$, which converges to $\mybar{\sigma}_{p,\infty}$ in the absence of an overload.

Further analytical justification of \eqref{steadystatedetector} is gained by
\begin{align*}
\Eset\left[\left.\widehat{\mybar{\sigma}}_{p,n}^2\right|\nooverloadnn\right]&\underset{\circledsmall{1}}{=}\frac{1}{n}\sum_{k=1}^{n}\Eset\left[\left.\left(\widehat{\mybar{v}}_{k}-\widehat{\mybar{v}}_{k}^p\right)^2\right|\nooverloadnn\right]\\
&\underset{\circledsmall{2}}{=}\frac{1}{n}\sum_{k=1}^{n}\Eset\left[\left.\left(\mybar{v}_{k}-\widehat{\mybar{v}}_{k}^p\right)^2\right|\nooverloadnn\right]\\
&\underset{\circledsmall{3}}{=}\frac{1}{n}\sum_{k=1}^{n}\frac{1}{\alpha_k^2}\cdot\Eset\left[\left.\left(e_{k}^p\right)^2\right|\nooverloadnn\right]\\
&\xrightarrow[\circledsmall{4}]{n\to\infty}\frac{\widetilde{\sigma}^2_{p,\infty}}{\alpha_{\infty}^2},
\end{align*}
where we have used:
\begin{enumerate}[i.]
	\setlength\itemsep{0.2em}
	\item[\circledjblue{1}] Linearity of the expectation;
	\item[\circledjblue{2}] Under $\mathcal{E}_{\tiny \mybar{{\rm{OL}}}_k}$, $\widehat{v}_k=v_k \;\Longrightarrow\; \widehat{\mybar{v}}_k=\mybar{v}_k$;
	\item[\circledjblue{3}] By definition \eqref{normalizedv}, \eqref{vbarlinearest}, \eqref{linearpredictionerror}, $\mybar{v}_k-\widehat{\mybar{v}}_{k}^p=e_{k}^p/\alpha_k$; and
	\item[\circledjblue{4}] Under the same ideal conditions described at the outset of this subsection, $\exists N_{\alpha_{\infty}}\in\Nset^+: \alpha_{n}=\alpha_{\infty}, \forall n>N_{\alpha_{\infty}}$; See Appendix \ref{AppB_approximation_of_cond_expectation} for a comment regarding this analysis.
\end{enumerate}
Therefore, as the adaptive process unfolds, the condition $\widehat{\mybar{\sigma}}_{p,n}>\frac{M_n}{2\kappa}$, which is a practical proxy for the ideal (merely theoretical) condition $\widehat{\mybar{\sigma}}_{p,n}=\frac{M_n}{2\kappa}$, and is essentially the decision rule for detecting the limit resolution $\alpha_{\infty}$, becomes increasingly accurate. Although the ideal conditions hold only approximately in practice, as we show in Section \ref{sec:simulresults} via simulations, the steady state detector \eqref{steadystatedetector} works quite well and is fairly accurate.

\begin{algorithm*}[t]
	\KwIn{Signal: $\{\left.x(t)\right|_{t=nT_s}\}$, System Parameters: $R, \alpha_0, p, \widehat{\uh}_0, \kappa, L_s, N_s, \epsilon_{\mu}, \delta_{\alpha}, \beta$}
	\KwOut{$\{\widehat{x}_{n}\}$}
	\nl Set $\alpha_n=\alpha_0$, $C_{\alpha}=0$, and $\widehat{\uh}_{p+1}=\widehat{\uh}_0$, and accumulate $p+1$ sample $\{y_i\}_{i=1}^{p+1}$; \algorithmiccomment{Set initial parameters}\\
	\nl Compute $\{\widetilde{y}_i=v_i\}_{i=1}^{p}$ as in \eqref{degeneratemodulo2}, $\ubv_{p+1}$ as in \eqref{normalizedv}, and set $\widehat{\ubv}_{p+1}=\ubv_{p+1}$; \algorithmiccomment{First $p$ samples are unfolded}\\
	\nl Compute $\widehat{\sigma}_{\bar{v},p+1}^2$ as in \eqref{varianceestofvbar}, and set $\mu=\epsilon_{\mu}/\left(p\cdot\widehat{\sigma}_{\bar{v},p+1}^2\right)$ as in \eqref{muourproposition}; \algorithmiccomment{Compute the LMS learning rate}\\
	\nl Set $C_{\alpha}=0$ and $\mathbbm{1}_{M_{\infty}}=0$; \algorithmiccomment{$C_{\alpha}$: \# iterations after adapting $\alpha_{n}$, $\mathbbm{1}_{M_{\infty}}$: $M_{\infty}$ flag}\\
	\For{$n=p+1,p+2,\ldots$}
	{
		\nl $\widehat{v}_n,\,\widehat{v}_n^p=\text{BlindModUnfold}(y_n, \widehat{\ubv}_n, \widehat{\uh}_n, R)$;\\
		\nl Update $\widehat{\sigma}_{\bar{v},n}^2=\frac{n-2}{n-1}\cdot\widehat{\sigma}_{\bar{v},n-1}^2+\frac{1}{n-1}\cdot\widehat{\mybar{v}}_{n-1}^2$;\\
		\eIf{$\widehat{\mathbbm{1}}_{\mathcal{E}_{\tiny {\rm{OL}}_n}}=1$}{
			\nl	Reset $\alpha_{n+1}=\alpha_0$, $C_{\alpha}=0$ and $\widehat{\uh}_{n+1}=\frac{\alpha_0}{\alpha_{n}}\cdot\widehat{\uh}_{n}$\label{resetalpha}; \algorithmiccomment{Re-open  modulo range}\\
			\nl Accumulate $(p+1)$ new samples $\{y_i\}_{i=n+1}^{n+p+1}$, compute $\{\widetilde{y}_i=v_i\}_{i=n+1}^{n+p}$; \algorithmiccomment{Re-initialization}\\
			\nl Output the respective estimates $\{\widehat{x}_{i}=\left(v_i+\frac{1}{2}\right)/\alpha_{0}\}_{i=n+1}^{n+p}$ of the accumulated samples;\\
			\nl Continue from $n=n+p+1$;
		}{
			\nl Output $\widehat{x}_n=(\widehat{v}_n+\frac{1}{2})/\alpha_{n}$;\\
			\nl Compute the estimated error $\widehat{e}_n^{\,p}=\widehat{v}_n-\widehat{v}^p_n$, and the estimated MSE $\widehat{\sigma}^2_{p,n}$ as in \eqref{eststdoflinearest};\\
			\nl Update $\widehat{\uh}_{n+1}=\widehat{\uh}_{n}+\mu\cdot\widehat{\ubv}_n\widehat{e}_n^{\,p}$, and increase $C_{\alpha}=C_{\alpha}+1$; \algorithmiccomment{LMS learning step}\\
			\If{$C_{\alpha}>L_s \cap \mathbbm{1}_{M_{\infty}}=0$}{
				\nl $\alpha_{n+1}=\left[\mathbbm{1}_{\uparrow\alpha,n}\cdot\frac{1}{\delta_{\alpha}}+\left(\mathbbm{1}_{\uparrow\alpha,n}-1\right)\cdot\delta_{\alpha}\right]\alpha_{n}$\label{alphaupdate}, $C_{\alpha}=0$; \algorithmiccomment{Update the resolution}\\ \vspace{0.05cm}
				\nl $\widehat{\uh}_{n+1}=\left[\mathbbm{1}_{\uparrow\alpha,n}\cdot\frac{1}{\delta_{\alpha}}+\left(\mathbbm{1}_{\uparrow\alpha,n}-1\right)\cdot\delta_{\alpha}\right]\widehat{\uh}_{n}$; \algorithmiccomment{Adapt the filter accordingly}\\
				\nl \textbf{if} $\widehat{\mathbbm{1}}_{M_{\infty},n}=1 \cap \mathbbm{1}_{\uparrow\alpha,n}=1$ \textbf{then} $\mathbbm{1}_{M_{\infty}}=1$\label{fixalphasteadystate}; \algorithmiccomment{Steady state detection};
			}
		}
	}
	\caption{{\bf Blind Modulo ADC Encoding-Decoding} \label{Algorithm3}}
\end{algorithm*}	
\setlength{\textfloatsep}{2pt}

It is also instructive to write the asymptotic resolution $\alpha_{\infty}$, via \eqref{equilibriumwithalpha} and $\Delta=2^R$, as,
\begin{equation}\label{asymptoticalpha}
\alpha_{\infty}=\frac{1}{\kappa}\cdot\left(\frac{1}{2}2^R\right)\cdot\frac{1}{\mybar{\sigma}_{p,\infty}}.
\end{equation}
The form \eqref{asymptoticalpha} provides several observations. First, and most obviously, increasing the number of bits $R$ increases the asymptotic resolution. Second, the trade-off in choosing the confidence level parameter $\kappa$ is now apparent. Indeed, increasing $\kappa$ leads to an exponential decrease in overload probability \eqref{boundgaussianQfunc}, but at the same time decreases the asymptotic resolution \eqref{asymptoticalpha}. Third, the inverse RMSE $1/\mybar{\sigma}_{p,\infty}$ reflects the \emph{unknown} causal and linear predictability accuracy. That is, how accurately the current sample of $\mybar{v}_n$ can be estimated, using a linear causal estimator, based on the previous $p$ samples $\ubv_n$. The lower $\mybar{\sigma}_{p,\infty}$, the higher the predictability, and accordingly, the higher the asymptotic resolution $\alpha_{\infty}$.

Interestingly, \eqref{asymptoticalpha} also provides a fresh look at the result \eqref{upperbound1} from \cite{ordentlich2018modulo}. Indeed, if we assume $\widetilde{\sigma}^2_{p,\infty}=\sigma^2_{\text{\tiny LMMSE},p}$, then using $\sigma_{\text{\tiny LMMSE},p}=\alpha_{\infty}\cdot\mybar{\sigma}_{p,\infty}$ from \eqref{LMMSEasalpha}, combined with \eqref{asymptoticalpha} written as $M_{\infty}=2\kappa\mybar{\sigma}_{p,\infty}$, the bound \eqref{upperbound1} reads,
\begin{equation}\label{consistentwithinformed}
\begin{aligned}
\Pr\left(\mathcal{E}^*_{\tiny {\rm{OL}}_n}\right)&\leq 2\exp\left\{-\frac{3}{2}2^{2\left(R-\frac{1}{2}\log_2(12\sigma_{\text{\tiny LMMSE},p}^2)\right)}\right\}\\
&=2\exp\left\{-\frac{3}{2}2^{2\left(\log_2(2^R)-\log_2(\sqrt{12}\alpha_{\infty}\cdot\bar{\sigma}_{p,\infty})\right)}\right\}\\
&=2\exp\left\{-\frac{3}{2}2^{2\log_2\left(\frac{M_{\infty}}{\sqrt{12}\bar{\sigma}_{p,\infty}}\right)}\right\}\\
&=2\exp\left\{-\frac{3}{2}2^{2\log_2\left(\frac{\kappa}{\sqrt{3}}\right)}\right\}=2\exp\left\{-\frac{\kappa^2}{2}\right\}.
\end{aligned}
\end{equation}
That is, for the ideal case in which the filter converges \emph{exactly} to the optimal one, i.e., the blind mod-ADC coincides with the oracle mod-ADC, the overload probability decreases exponentially with $\kappa^2$. This is in perfect compliance with \eqref{boundgaussianQfunc}.

Yet another way to see the consistency of the blind mod-ADC asymptotic performance with the that of the informed mod-ADC is via the asymptotic rate. By isolating $R$ in \eqref{asymptoticalpha}, using \eqref{LMMSEasalpha}, and $\widetilde{\sigma}^2_{p,\infty}=\sigma^2_{\text{\tiny LMMSE},p}$ as above, we have
\begin{align}
R&=\frac{1}{2}\log_2\left(\frac{\mybar{\sigma}^2_{p,\infty}}{\frac{1}{12\alpha^2_{\infty}}}\right)+\log_2\left(\frac{\kappa}{\sqrt{3}}\right)\nonumber\\
    &\triangleq\frac{1}{2}\log_2\left(12\sigma^2_{\text{\tiny LMMSE},p}\right)+\delta_{\kappa},\label{asyptoticrate}
\end{align}
which, again, is in perfect compliance with \cite{ordentlich2018modulo}, Eq.\ 12 therein, such that $\kappa$ controls the overload probability, and inevitably the excess rate $\delta_{\kappa}$ with respect to Shannon's lower bound \cite{berger1971ratedistortion}.

Having described in detail all the individual components, namely initialization, resolution adaptation, error propagation prevention, and steady state detection, we are now ready to present the complete algorithm of the blind mod-ADC encoder-decoder, given in Algorithm \ref{Algorithm3}.

An important observation is that our algorithm can work \emph{without} incorporating the steady state detector. In other words, \eqref{steadystatedetector} is not a necessary component required in order to ensure proper operation of the blind mod-ADC. Moreover, in some cases, when working in highly dynamic environments, we might intentionally choose to disable this detector, thus allowing the LMS to continuously adapt the linear filter $\widehat{\uh}_n$ according to the input, whose SOSs may change over time. 

Nevertheless, in a broader view of the proposed architecture, bare in mind that the primary cost paid for the desired robustness is the addition of a Digital Signal Processing (DSP) unit, running the adaptive algorithm (see Fig.\ \ref{fig:Blind_mod_ADC_block_diagram}). Therefore, whenever possible, reducing the operation time of this (extra) DSP unit (relative to the informed mod-ADC), reduces the power consumption of the entire mod-ADC as a whole. We conclude that, whenever desired, if $\widehat{\mathbbm{1}}_{M_{\infty},n}$ is \emph{fixed} to $0$, the algorithm continues to work properly, and is able to track dynamics reflected in the SOSs of the input. The only difference in performance would be that, for a stationary input, the asymptotic resolution will oscillate around \eqref{asymptoticalpha}, since the LMS only converges in mean, and not to a fixed filter. 

Lastly, this notion naturally extends to non-stationary input signals. In these case, if the associated SOSs vary sufficiently slow (relative to the sampling period $T_s$), the LMS and its triggering estimator \eqref{eststdoflinearest} are constantly fed with ``quasi-stationary" inputs, the resolution is constantly being adapted, and the blind mod-ADC continues to work properly.

\section{Simulation Results}\label{sec:simulresults}
In this section, we present empirical results of two simulation experiments, which demonstrate the successful operation of our proposed algorithmic framework. These results corroborate our analytical derivations in Section \ref{sec:proposedsolution}, and to the best of our knowledge, serve as the first empirical evidence for the implementation feasibility of a blind mod-ADC for scalar time series input signals.

\subsection{Experiment 1: A Gaussian Input Signal}\label{subsec:Gaussiansignal}
\begin{table}
	\centering
	{\renewcommand{\arraystretch}{1.5}
		\begin{tabular}{||c c||} 
			\hline
			\textbf{Parameter} & \textbf{Value} \\ [0.5ex] 
			\hline\hline
			$R\;(\rightarrow \Delta=2^R)$ & $10\;(\rightarrow 1024)$  \\ [0.25ex]
			\hline
			$\alpha_{0}\;(\rightarrow M_0=2^R/\alpha_{0})$ & $100\;(\rightarrow 10.24)$  \\ [0.25ex]
			\hline
			$p$ & $40$ \\ [0.25ex]
			\hline
			$\widehat{\uh}_{0}$ & $\left[1\;0\;\cdots\;0\right]^{\tps}\in\Rset^{p\times 1}$ \\ [0.25ex]
			\hline
			$\kappa$ & $4.5$ \\ [0.25ex]
			\hline
			$L_s$ & $p=40$ \\
			\hline
			$N_s$ & $500$ \\
			\hline
			$\epsilon_{\mu}$ & $10^{-2}$\\
			\hline
			$\delta_{\alpha}$ & $0.9$\\
			\hline
			$\beta$ & $5$\\
			\hline
		\end{tabular}
		\caption{Chosen system parameters for simulation experiment 1.}
		\label{table:inputsystemparams}
	}
\end{table}
\begin{figure*}
	\centering
	\begin{subfigure}{0.45\textwidth}
		\centering
		\includegraphics[width=\textwidth]{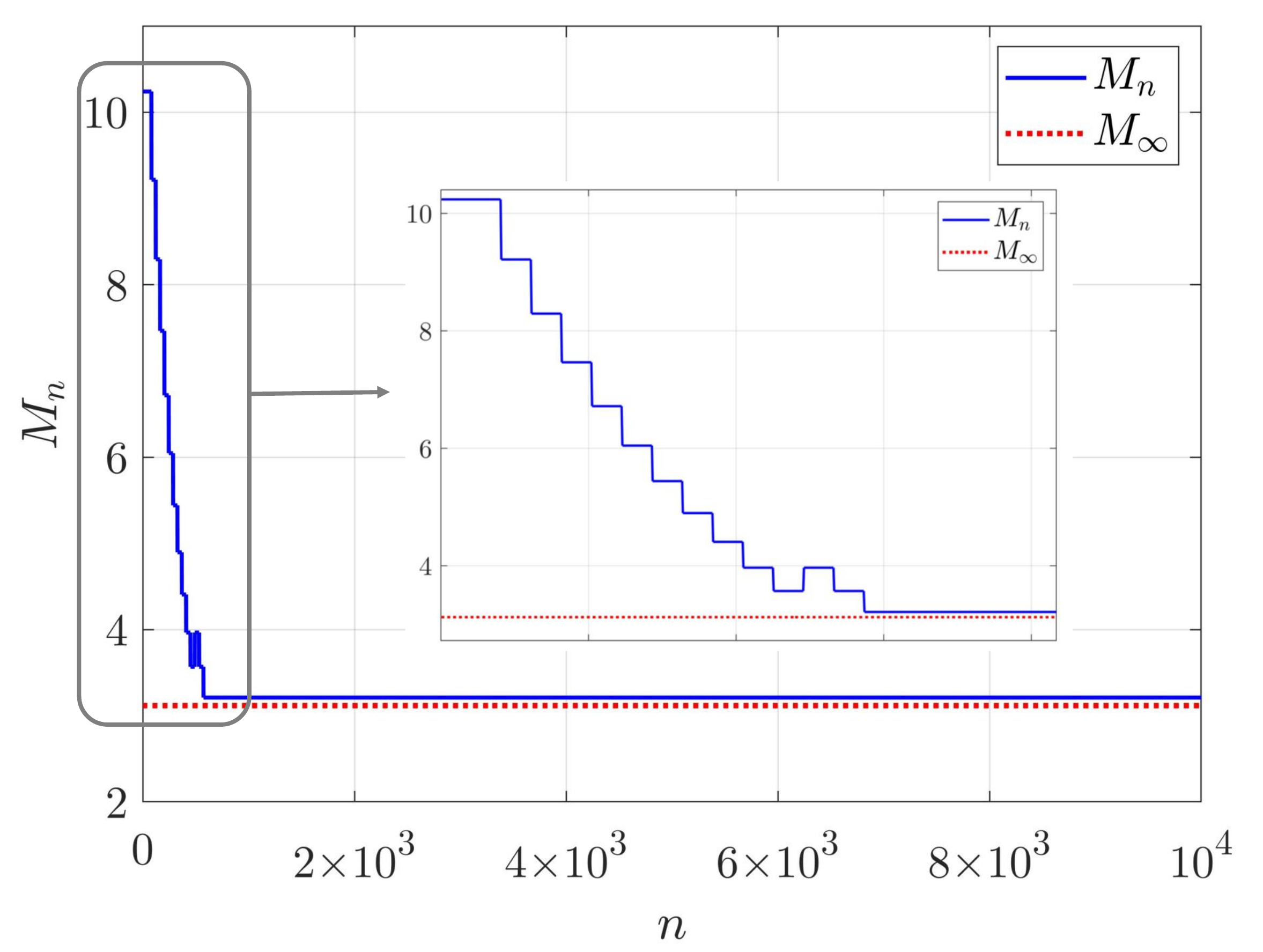}\vspace{-0.2cm}
		\caption{}
		\label{fig:effectivemodconvergencesimul1}
	\end{subfigure}
	\begin{subfigure}{0.48\textwidth}
		\centering
		\includegraphics[width=\textwidth]{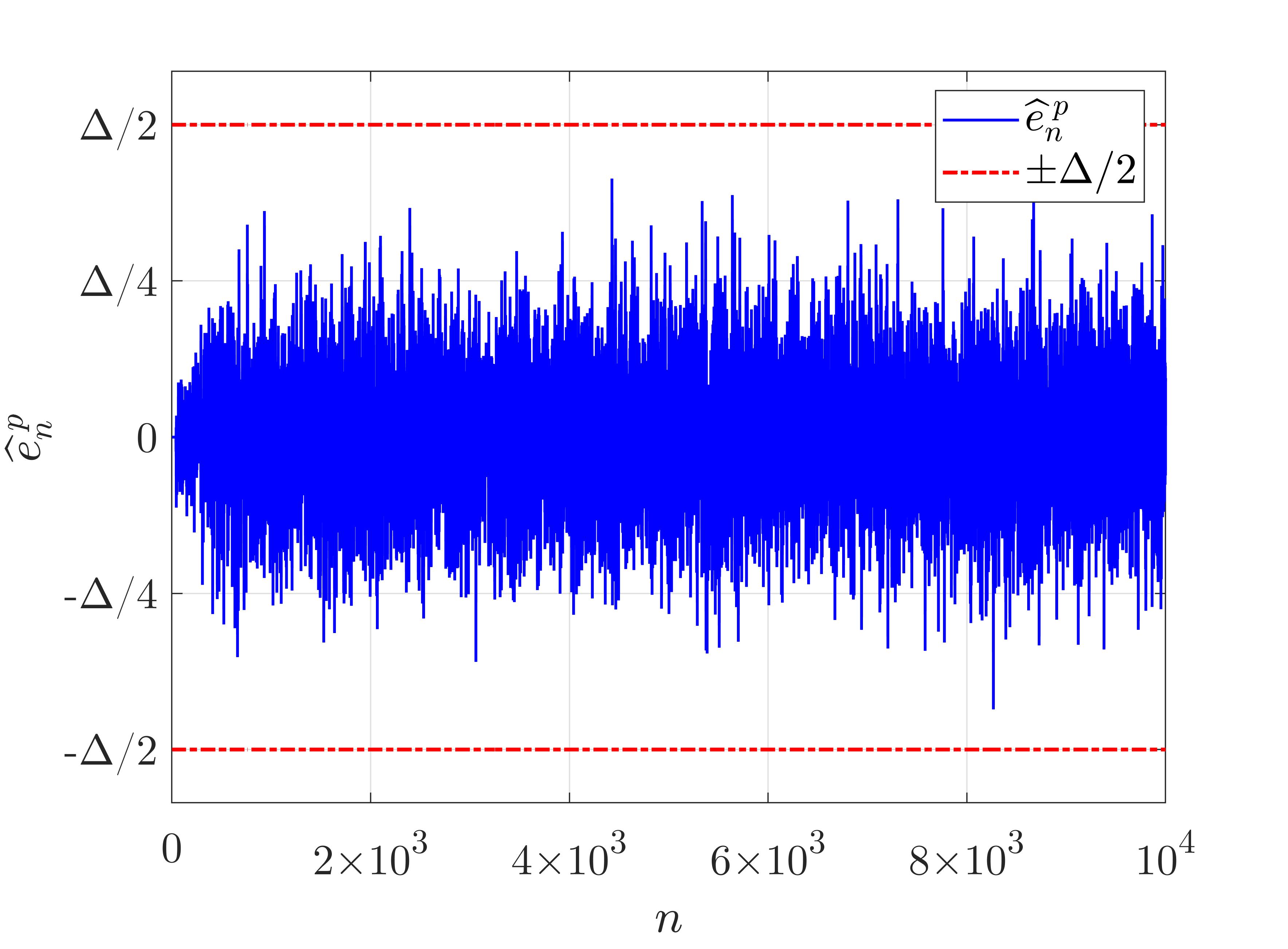}\vspace{-0.2cm}
		\caption{}
		\label{fig:estimatederrorsimul1}
	\end{subfigure}
	\caption{Results of simulation experiment 1, with $\kappa=4.5$. (a) The effective modulo range $M_n$ vs.\ discrete-time. Clearly, since on this time interval an overload did not occur, the resolution $\alpha_{n}$ approximately converged to the asymptotic resolution $\alpha_{\infty}$, as predicted by our analytical asymptotic analysis. Upon convergence, stability is also evident, and this is due to the successful operation of the steady state detector \eqref{steadystatedetector}. (b) The estimated errors process $\widehat{e}_n^{\,p}$ vs.\ discrete-time. It is seen that $|\widehat{e}_n^{\,p}|<\frac{1}{2}2^R$ for the entire time interval. Since here $\widehat{e}_n^{\,p}=e_n^p$, an overload event $\mathcal{E}_{\tiny {\rm{OL}}_n}$ did not occur, and we obtain perfect recovery of the process $v_n$.}
	\label{fig:figureexp1main}
\end{figure*}
\begin{figure*}
	\centering
	\begin{subfigure}{0.48\textwidth}
		\centering
		\includegraphics[width=\textwidth]{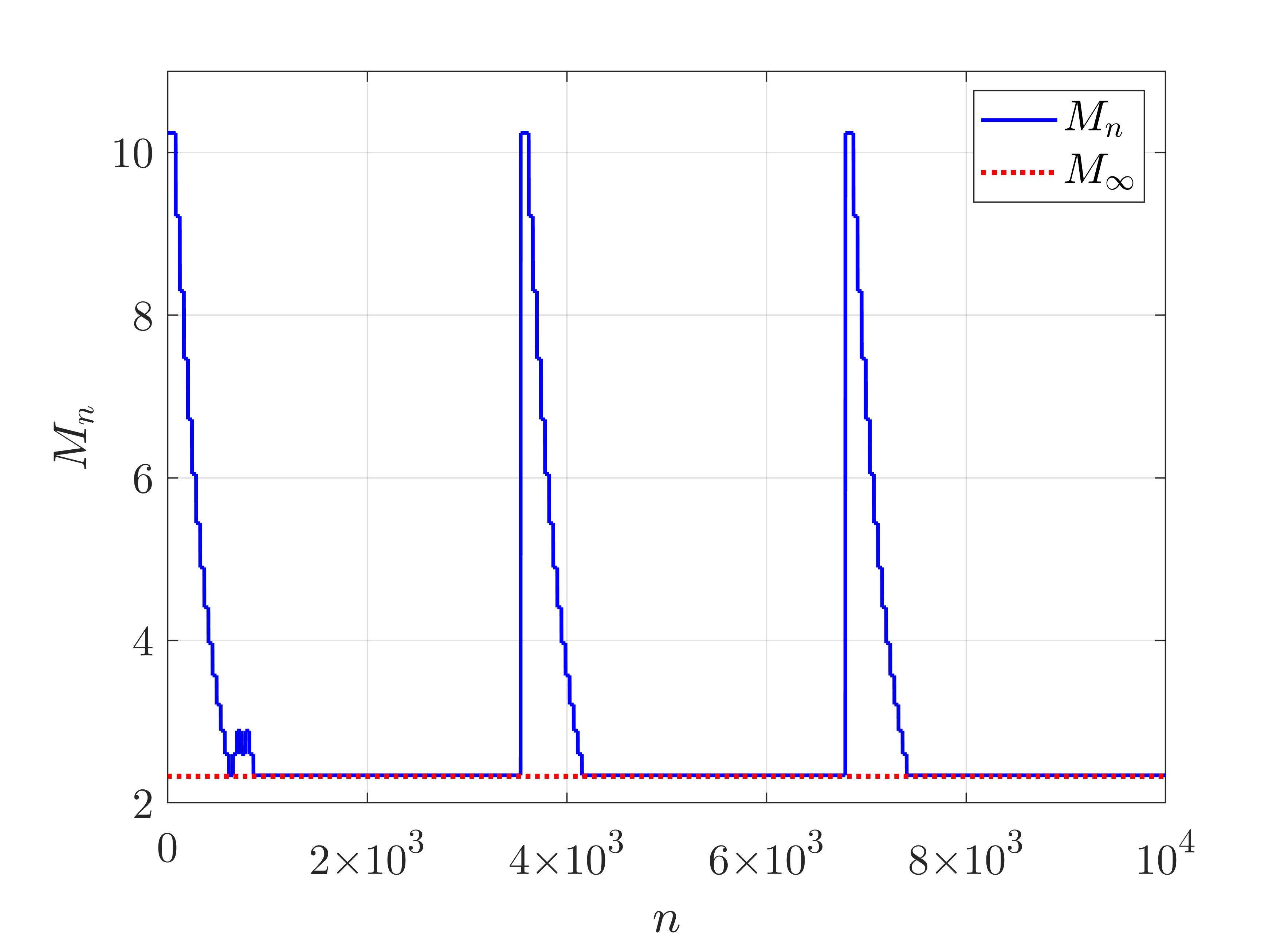}\vspace{-0.2cm}
		\caption{}
		\label{fig:exp1fig2kappasmaller}
	\end{subfigure}
	\begin{subfigure}{0.48\textwidth}
		\centering
		\includegraphics[width=\textwidth]{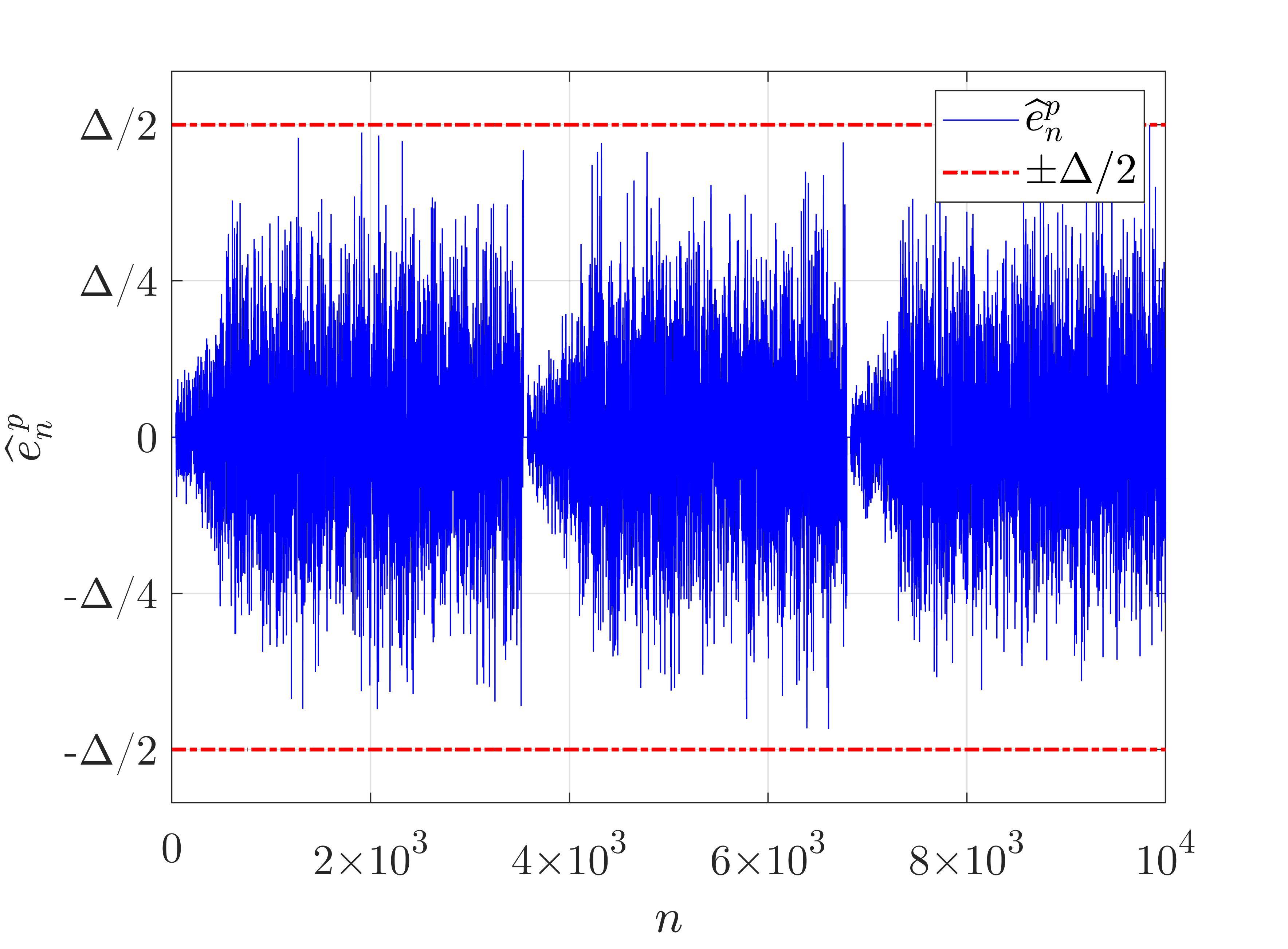}\vspace{-0.2cm}
		\caption{}
		\label{fig:exp1fig3kappasmaller}
	\end{subfigure}
	\caption{Results of simulation experiment 1, but with $\kappa=3.5$ instead of $\kappa=4.5$. (a) The effective modulo range $M_n$ (b) The estimated errors process $\widehat{e}_n^{\,p}$, both vs.\ the discrete-time index $n$. For a lower value of $\kappa$, overload events are more frequent. Nonetheless, our algorithm successfully detects these events, and automatically lowers the resolution in order to maintain proper continuous operation of the blind mod-ADC.}
	\label{fig:samefigurewithkappasmaller}
\end{figure*}
We consider the case where the input signal $x_n$ is Gaussian. This is quite a common assumption; for example, digital communication signals are commonly modeled as Gaussian, see, e.g., \cite{banelli2000theoretical}. Specifically, we generate the input as
\begin{equation}\label{inputMAprocess}
x_n = \frac{1}{\sqrt{L_x}}\sum_{\ell=0}^{L_x-1}\xi_{n-\ell},
\end{equation}
where $\{\xi_n\}$ is a zero-mean unit-variance Gaussian i.i.d.\ process. Accordingly, the autocorrelation function of $x_n$, assumed to be unknown, is given by
\begin{equation}\label{autocorrelationofinoutexp1}
R_x[\ell] = \left(1-\frac{|\ell|}{L_x}\right)\cdot\mathbbm{1}_{|\ell|<L_x},
\end{equation}
such that $\sigma_x^2=1$, i.e., $x_n$ is also zero-mean and unit-variance. Notice that $L_x$ is the one-sided support of the autocorrelation function $R_x[\ell]$. In particular, this parameter directly affects the (unknown) level of predictability $1/\mybar{\sigma}_{p,\infty}$, which appears in \eqref{asymptoticalpha}, and therefore implicitly determines the resolution $\alpha_{\infty}$.

We simulate a quantizer with $R=10$ bits, and generate the signal $v_n$ according to \eqref{defofv}. We then apply the $2^R$-modulo operator to $v_n$, which yields the simulated mod-ADC output process $y_n$, as in \eqref{outputofmodADC}. The chosen set of required system parameters, prescribed in the input to Algorithm \ref{Algorithm3}, is given in Table \ref{table:inputsystemparams}. Further, we set $L_x=15$, and emphasize that we intentionally choose $p\neq L_x$, and specifically $p>L_x$. This simulates the more probable scenario, in which the support of $R_x[\ell]$ is unknown, hence the length of the FIR filter $\widehat{\uh}_n$ will not be perfectly matched to the length of the optimal LMMSE causal filter, which is of length $L_x-1$ in this case. Moreover, we choose $L_s=p$ in order to demonstrate that calibration of the system parameters could be simple, and rather straightforward. Note also that we choose $\kappa=4.5$, which gives $\Pr\left(\left.|e_n^p|>\frac{1}{2}2^R\right|\nooverloadnp\right)=2Q\left(4.5\right)\approx6.7953\times10^{-6}$.

To demonstrate a typical operation of the proposed blind mod-ADC, we first consider a realization of length $N=10^4$ samples. Fig.\ \ref{fig:effectivemodconvergencesimul1} presents the effective modulo range $M_n$ \eqref{effectivemodulorange} vs.\ the discrete-time index $n$. Starting from $M_0=10.24$, more than $10$ times the standard deviation of the input $x_n$, effectively guarantees that $v_n=\widetilde{y}_n$, i.e., no folding occurs during the initialization phase, as desired. This provides the necessary conditions for the LMS algorithm to learn the optimal filter. When $C_{\alpha}>L_s$, namely, a resolution update is allowed, $\alpha_{n}$ is increased, as can be seen more conveniently in superimposed ``close-up" of the convergence interval. 

The adaptive process continues with updates of multiplicative step-sizes $\delta_{\alpha}$, and whenever required, is also decreased. Furthermore, it is seen that at some point, the steady state detector $\widehat{\mathbbm{1}}_{M_{\infty},n}$ is turned on, indicating that the asymptotic resolution has been approximately attained. Indeed, the convergence is not exactly to $M_{\infty}$ (equivalently to $\alpha_{\infty}$), since in practice $\widehat{\sigma}^2_{p,n}\neq\sigma^2_{p,n}$ with probability $1$ and the adaptations of $\alpha_{n}$ are of finite resolution. Nonetheless, as evident from Fig.\ \ref{fig:effectivemodconvergencesimul1}, the optimistic asymptotic analysis, carried out under ideal theoretical conditions, provides a considerably accurate forecast of the steady state resolution.

The \emph{estimated} estimation errors $\widehat{e}_n^{\,p}$ of the linear estimator $\widehat{v}_n^p$ are presented in Fig.\ \ref{fig:estimatederrorsimul1}. Recall that in order to unfold $\widetilde{y}_n$, these estimation error \eqref{estimatedestimationerrors} are necessary, and $\widehat{e}_n^{\,p}=e_n^p$ hold only when there is no overload. As reflected from Fig.\ \ref{fig:estimatederrorsimul1}, this is exactly the case, since $\mathcal{E}_{\tiny \mybar{{\rm{OL}}}_n}=\{\left|e_{n}^p\right|<\frac{1}{2}2^R\}=\left\{\widehat{e}_n^{\,p}=e_n^p\right\}$. Accordingly, in this experiment the averaged squared error is $\frac{1}{N}\sum_{n=1}^{N}\left(v_n-\widehat{v}_n\right)^2\approx5.067\times10^{-27}$, which is clearly due to machine accuracy limitations, thus implying perfect recovery of the signal $v_n$, from which $x_n$ can be readily estimated.

We repeat the experiment with exactly the same setting, only now with $\kappa=3.5$. In a typical operation, we expect to observe an increased asymptotic resolution, at the cost of a more frequent overload event. Figs.\ \ref{fig:exp1fig2kappasmaller} and \ref{fig:exp1fig3kappasmaller} presenting the corresponding plots as in Figs.\ \ref{fig:effectivemodconvergencesimul1} and \ref{fig:estimatederrorsimul1}, respectively, reflect exactly this trend. Nevertheless, in the case of an overload event, our error propagation defense mechanism comes into play, and maintains proper continuous operation, as it is well demonstrated in Fig.\ \ref{fig:exp1fig2kappasmaller}. Thus, the blind mod-ADC automatically balances the trade-off between effective quantization and continuous operation, which is highly important in practice. These results corroborate our analytical derivation of \eqref{ourdetector}.

\subsection{Experiment 2: A Bandlimited Input with Narrowband Interferers}\label{subsec:narrowbandinterference}
\begin{table}
	\centering
	{\renewcommand{\arraystretch}{1.5}
		\begin{tabular}{||c c c c||} 
			\hline
			\textbf{Parameter} & $i=1$ & $i=2$ & $i=3$ \\ [0.5ex] 
			\hline\hline
			$g_i$ & $2$ & $2$ & $2$  \\ [0.25ex]
			\hline
			$\omega_i$ & $\pi/4$ & $4\pi/5$ & $\sqrt{2}\pi/3$  \\ [0.25ex]
			\hline
			$\tau_i$ & $2\times10^4$ & $4\times10^4$ & $7\times10^4$  \\ [0.25ex]
			\hline
		\end{tabular}
		\caption{Chosen interference parameters for simulation experiment 2. Note that $\omega_3$ is not a rational multiplication of $\pi$, hence $\zeta^3_n$ is not a periodic signal.}
		\label{table:interefernceparams}
	}
\end{table}
\begin{figure}[t]
	\includegraphics[width=0.5\textwidth]{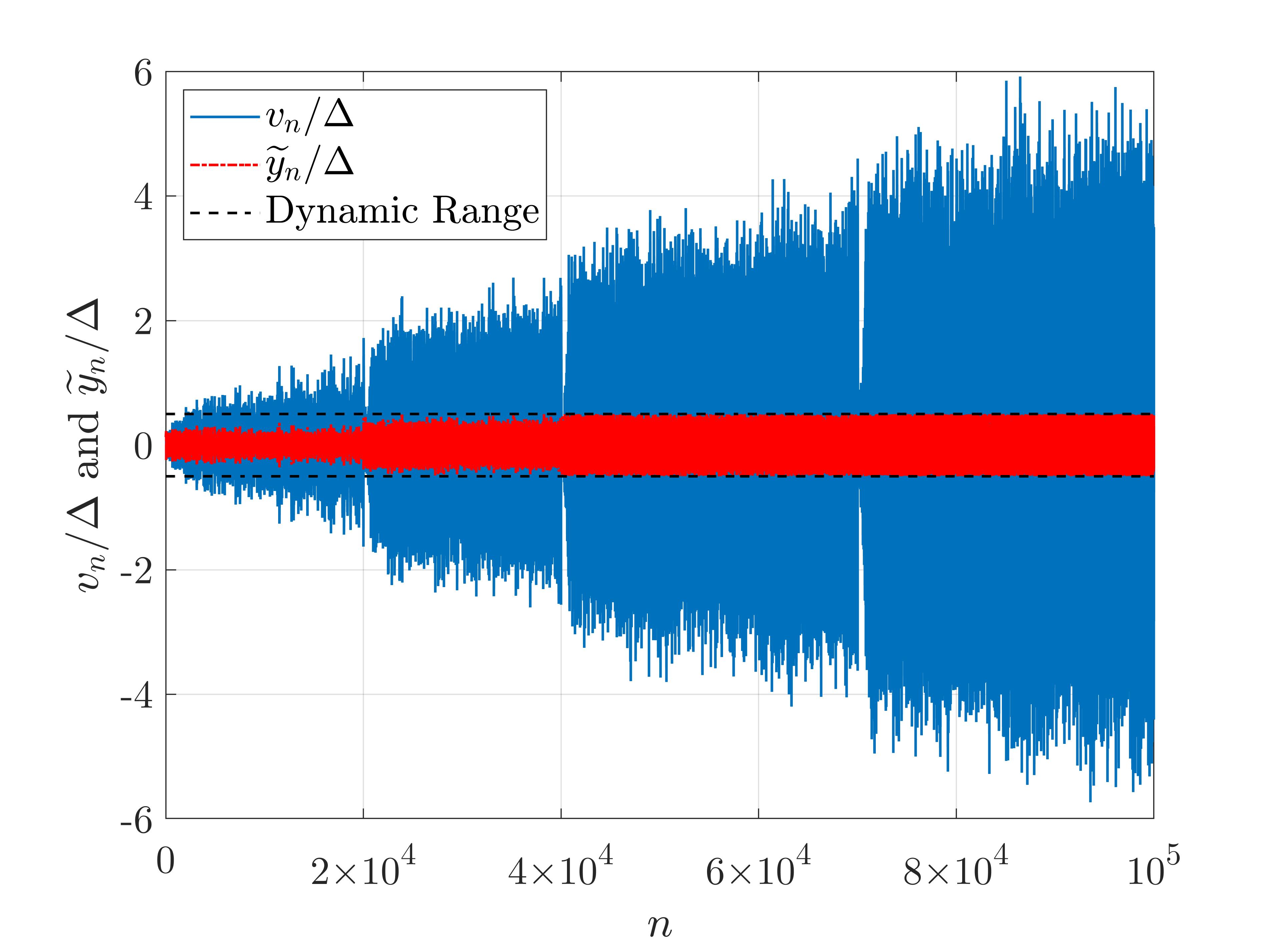}
	\centering
	\caption{Experiment 2: The unfolded and folded signals $v_n$ and $\widetilde{y}_n$, respectively, normalized by $\Delta=2^R$. In such a scenario, due to the interference signals, a standard ADC would have been saturated, most likely. In our case, the modulo folding is fully operational, and the unfolded signal can still be recovered.}
	\label{fig:unfoldedandfoldedsimul2}
\end{figure}
\begin{figure}
	\includegraphics[width=0.5\textwidth]{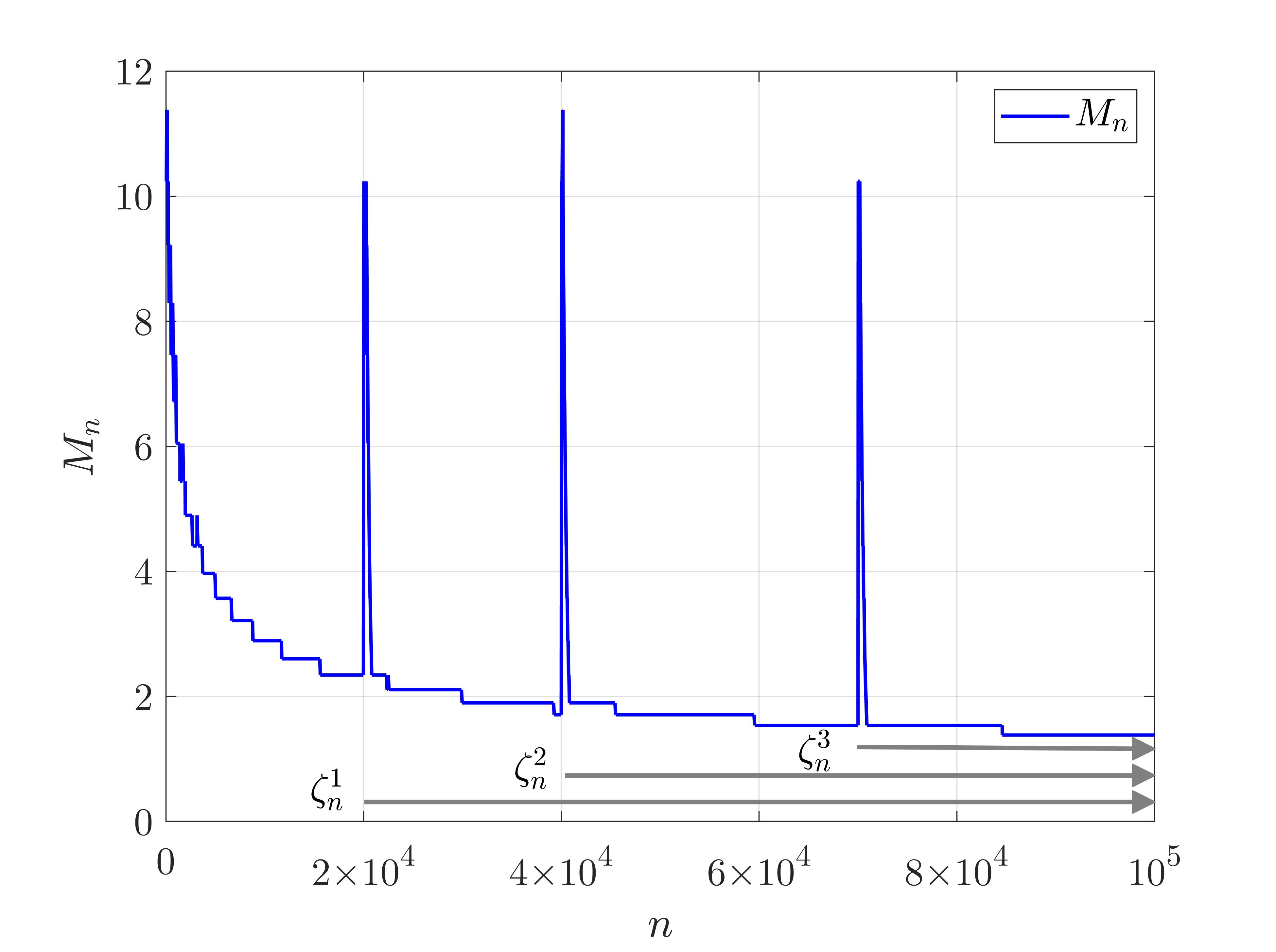}
	\centering
	\caption{The effective modulo range $M_n$ vs.\ discrete-time. It is seen that our adaptive algorithm quickly recovers from the abrupt change due to the interferences, re-learns the suitable filter, and returns to a high-resolution operational mode.}
	\label{fig:effectivemodconvergencesimul2}
\end{figure}
In this experiment we consider a non-Gaussian signal of interest, and the presence of narrowband interferences. Specifically, here the signal of interest $\widetilde{x}_n$ is generated by applying a non-ideal, minimum-order filter with a stopband attenuation of $60$ dB, to the driving noise $\{\xi_n\}$, which is drawn from the Rademacher distribution, namely $\Pr(\xi_n=1)=\Pr(\xi_n=-1)=\frac{1}{2}$. We then normalize the output, such that $\widetilde{x}_n$ is a zero-mean unit-variance process. In addition, we consider the presence of narrowband interference signals, a particularly relevant scenario in the context of communication systems.

Thus, the input to the blind mod-ADC in this experiment, which is of length $N=10^5$ samples, is given by
\begin{equation}\label{signalandintereference}
x_n = \widetilde{x}_n + \sum_{i=1}^{3}\underbrace{g_i\cdot\sin(\phi_i+\omega_in)\cdot\mathbbm{1}_{n>\tau_i}}_{\triangleq\zeta^i_n}=\widetilde{x}_n + \sum_{i=1}^{3}\zeta^i_n,
\end{equation}
where $\widetilde{x}_n$ simulates the signal of interest, and $\{\zeta^i_n\}_{i=1}^3$ simulate three narrowband interference signals. Here, for the interference $\zeta_i$, the parameters $g_i, \phi_i, \omega_i$ and $\tau_i$ are the unknown gain, phase, carrier (angular) frequency and transmission start time, respectively. We draw $\phi_i\sim{\rm{Unif}}\left((0,2\pi]\right)$ independently, and set the rest of the parameters as reported in Table \ref{table:interefernceparams}. We also fix $\widehat{\mathbbm{1}}_{M_{\infty},n}$ to zero, to demonstrate the unnecessity of the steady state detector for a successful operation.

The spectrogram of the input signal $x_n$ is presented in Fig.\ \ref{fig:introfigurea}. During the starting period ($n<2\times10^4$), the blind mod-ADC needs to operate in the presence of the signal of interest $\widetilde{x}_n$ only. As soon as $n>\tau_1$, the first interference is added, and the scenario becomes even harder after $\tau_2$ and $\tau_3$, where the signal of interest is impaired by the interferes. This is easily seen from Fig.\ \ref{fig:unfoldedandfoldedsimul2}, presenting the unfolded and folded signals, $v_n$ and $\widetilde{y}_n$, respectively.

Fig.\ \ref{fig:effectivemodconvergencesimul2} presents the evolution of the effective modulo range $M_n$ in (discrete) time. Whenever an interference starts transmitting, the SOSs of the input signal abruptly changes, and as a result the estimation error $e_n^p$ increases dramatically, thus causing an overload event. We recall that an overload, by definition (see \eqref{defOLofaaptive}), \emph{does not} mean that the amplitude of the input signal exceeds the modulo range $\Delta=2^R$, as in a standard ADC. Rather, an overload event occurs when the magnitude of the estimation error $e_n^p$ exceeds half the modulo range. In these cases, our overload detector detects these large estimation errors, and the effective range is re-opened. This way, a stream of low-resolution, though \emph{unfolded} samples are produced, allowing the LMS algorithm to re-learn the new (and different) optimal filter. This process happens right after $\tau_1, \tau_2$ and $\tau_3$, but operation in high-resolution is gained anew. 

We also report the average number of failures in perfectly recovering $v_n$, i.e., the empirical error probability, which is
\begin{equation*}
\widehat{\Pr}(v_n\neq\widehat{v}_n)\triangleq\frac{1}{N}\sum_{\ell=1}^{N}\mathbbm{1}_{v_n\neq\hat{v}_n}=7.8\cdot10^{-4}=0.078\%.
\end{equation*}
Evidently, the blind mod-ADC provides highly reliable recovery of the unfolded signal, and in turn, allows for highly accurate estimation of the input $x_n$ via \eqref{estimateofx}. In this specific scenario, since the input \eqref{signalandintereference} is perfectly recovered (almost everywhere), the narrowband interferers can be easily detected and filtered out (e.g., using notch filters). This digital solution, which is now simple thanks to the modulo-ADC, could not have been achieved by a standard, probably saturated ADC.

We emphasize that all the results reported in this Section were verified by multiple runs, and were consistently observed for multiple (different) realizations.

\section{Conclusion}\label{sec:conclusion}
In the context of analog-to-digital conversion, we have presented an algorithmic framework, allowing for a stable and reliable operation of a mod-ADC without access to prior knowledge of the input signal's SOSs. We put forth the key design parameters, and discussed the corresponding trade-offs. In addition, we derived the asymptotic resolution of the proposed blind mod-ADC, and linked our current result with the performance of the previously presented oracle mod-ADC \cite{ordentlich2018modulo}. We demonstrated by simulations the successful operation of our proposed solution, which corroborated its underlying theoretical infrastructure. Moreover, we demonstrated the advantage in using a mod-ADC in an environment of multiple interference signals.

As ADCs are being used in a host of applications, more often than not when perfect knowledge on the input signal is not available (if at all), the robustness of such devices is imperative, and almost crucial. The ability of operating blindly under dynamic conditions is essential for practical purposes, and constitutes a key advantage in effective sensing. Therefore, this work is yet another important step towards realization of mod-ADCs, shrinking the gap between sensing performance in practice and the respective theoretical limits.

\appendices

\section{SOSs of the Processes $v_n$ and $\mybar{v}_n$}\label{AppA_vbar_vs_v}
As explained in Subsection \ref{subsec:phase1initialization}, when $\alpha_n$ increases, the SOSs of $\{\mybar{v}_n\}$ gradually become less affected by $\alpha_n$, which is not true for $\{v_n\}$. To see this more clearly, observe that the autocorrelation function of $\{\mybar{v}_n\}$ is given by
\begin{equation}
R_{\bar{v}}[\ell]\triangleq\Eset\left[\mybar{v}_n\mybar{v}_{n-\ell}\right]=R_x[\ell]+\mathbbm{1}_{\ell=0}\cdot\frac{1}{12\alpha_n^2},
\end{equation}
where we have used (i) the definition \eqref{normalizedv}; (ii) the fact that $x_n$ and $z_n$ are statistically independent; and (iii) $z_n$ is an i.i.d.\ process. In contrast, the autocovariance of $\{v_n\}$ is given by
\begin{equation}
\begin{aligned}
R_{v}[\ell]&\triangleq\Eset\left[\left(v_n-\frac{1}{2}\right)\left(v_{n-\ell}-\frac{1}{2}\right)\right]\\
&=\alpha_n\alpha_{n-\ell}R_x[\ell]+\mathbbm{1}_{\ell=0}\cdot\frac{1}{12},
\end{aligned}
\end{equation}
and we examine the autocovariance (rather than the autocorrelation) since $\Eset[v_n]=\frac{1}{2}$. 

Evidently, for any non-zero lag $\ell\neq0$, $R_{\bar{v}}[\ell]=R_x[\ell]$, and in particular $R_{\bar{v}}[\ell]$ is independent of $\alpha_n$ for $\ell\in\{1,\ldots,p\}$. Clearly, this is not the case for $R_{v}[\ell]$. Moreover, the variance $R_{\bar{v}}[0]$, also given in \eqref{powerofvtilde}, approaches $\sigma_x^2$ as $\alpha_n$ increases. Since $R_{\bar{v}}[\ell]$ is less sensitive than $R_v[\ell]$ to adaptations in $\alpha_n$, by using the recovered values of $\{\mybar{v}_n\}$ rather than $\{v_n\}$ as the observations in \eqref{linearestimation} (as opposed to \eqref{LMMSElengthp}), we alleviate the estimation (/learning) of the optimal filter coefficients, which depend on the SOSs of the observations, throughout the adaptive process.

\section{Comment on the Asymptotic RMSE Estimator}\label{AppB_approximation_of_cond_expectation}
Although $\widehat{\mybar{\sigma}}_{p,n}^2$, defined in \eqref{sigmabarest}, is perhaps the most intuitive estimator of $\mybar{\sigma}_{p,n}^2$, an exact analysis of its asymptotic properties is far from trivial. Indeed, since it is a random process whose statistical properties are implicitly determined by the resolution update, error propagation prevention and steady state detection rules (steps \ref{alphaupdate}, \ref{resetalpha} and \ref{fixalphasteadystate} in Algorithm \ref{Algorithm3}, respectively), it is even non-stationary to begin with.

However, to further justify our proposed steady state detector \eqref{steadystatedetector}, which is based on $\widehat{\mybar{\sigma}}_{p,n}^2$ \eqref{sigmabarest}, it suffices to consider a simplified scenario, in which an overload event never occurs. Of course, this happens with probability zero when considering an infinitely long observation of the error process $\{e_n^p\}$, since we assume $\left.e_n^p\right|\mathcal{E}_{\tiny \mybar{{\rm{OL}}}_{n}}^{(p)}\sim\mathcal{N}(0,\sigma_{p,n}^2)$. Nevertheless, such an analysis is informative for finite, but sufficiently long realizations, in which our proposed adaptive mechanism for the blind mod-ADC converges to steady state, i.e., $\widehat{\mathbbm{1}}_{M_{\infty},n}=1$, which occurs w.h.p.\ with proper selection of the system parameters. The simulation results presented in Section \ref{sec:simulresults} corroborate this argument, and further justifies this approach for a simplified, yet informative analysis, resulting in \eqref{asymptoticalpha}, which is consistent with the analysis of the informed mod-ADC presented in \cite{ordentlich2018modulo}, as evident from \eqref{consistentwithinformed}.

\bibliography{Bibfile}

\begin{thebibliography}{10}
\providecommand{\url}[1]{#1}
\csname url@samestyle\endcsname
\providecommand{\newblock}{\relax}
\providecommand{\bibinfo}[2]{#2}
\providecommand{\BIBentrySTDinterwordspacing}{\spaceskip=0pt\relax}
\providecommand{\BIBentryALTinterwordstretchfactor}{4}
\providecommand{\BIBentryALTinterwordspacing}{\spaceskip=\fontdimen2\font plus
\BIBentryALTinterwordstretchfactor\fontdimen3\font minus
  \fontdimen4\font\relax}
\providecommand{\BIBforeignlanguage}[2]{{%
\expandafter\ifx\csname l@#1\endcsname\relax
\typeout{** WARNING: IEEEtran.bst: No hyphenation pattern has been}%
\typeout{** loaded for the language `#1'. Using the pattern for}%
\typeout{** the default language instead.}%
\else
\language=\csname l@#1\endcsname
\fi
#2}}
\providecommand{\BIBdecl}{\relax}
\BIBdecl

\bibitem{wang2010advances}
B.~Wang and K.~R. Liu, ``Advances in cognitive radio networks: A survey,''
  \emph{IEEE Journal of selected topics in signal processing}, vol.~5, no.~1,
  pp. 5--23, 2010.

\bibitem{cabric2005physical}
D.~Cabric and R.~W. Brodersen, ``Physical layer design issues unique to
  cognitive radio systems,'' in \emph{2005 IEEE 16th International Symposium on
  Personal, Indoor and Mobile Radio Communications}, vol.~2, 2005, pp.
  759--763.

\bibitem{ordentlich2018modulo}
O.~Ordentlich, G.~Tabak, P.~K. Hanumolu, A.~C. Singer, and G.~W. Wornell, ``A
  modulo-based architecture for analog-to-digital conversion,'' \emph{IEEE
  Journal of Selected Topics in Signal Processing}, vol.~12, no.~5, pp.
  825--840, 2018.

\bibitem{sun2010automatic}
F.~Sun, J.~Singh, and U.~Madhow, ``Automatic gain control for {ADC}-limited
  communication,'' in \emph{2010 IEEE Global Telecommunications Conference
  GLOBECOM 2010}, 2010, pp. 1--5.

\bibitem{er79}
T.~Ericson and V.~Ramamoorthy, ``Modulo-{PCM}: A new source coding scheme,'' in
  \emph{ICASSP '79. IEEE International Conference on Acoustics, Speech, and
  Signal Processing}, vol.~4, Apr 1979, pp. 419--422.

\bibitem{zamir2008achieving}
R.~Zamir, Y.~Kochman, and U.~Erez, ``Achieving the {G}aussian rate--distortion
  function by prediction,'' \emph{IEEE Transactions on Information Theory},
  vol.~54, no.~7, pp. 3354--3364, 2008.

\bibitem{oe17}
O.~Ordentlich and U.~Erez, ``Integer-forcing source coding,'' \emph{IEEE
  Transactions on Information Theory}, vol.~63, no.~2, pp. 1253--1269, Feb
  2017.

\bibitem{romanov2021blind}
E.~Romanov and O.~Ordentlich, ``Blind unwrapping of modulo reduced {G}aussian
  vectors: Recovering {MSB}s from {LSB}s,'' \emph{IEEE Transactions on
  Information Theory}, vol.~67, no.~3, pp. 1897--1919, 2021.

\bibitem{bhandari2017unlimited}
A.~Bhandari, F.~Krahmer, and R.~Raskar, ``On unlimited sampling,'' in
  \emph{2017 International Conference on Sampling Theory and Applications
  (SampTA)}.\hskip 1em plus 0.5em minus 0.4em\relax IEEE, 2017, pp. 31--35.

\bibitem{bkr18isit}
------, ``Unlimited sampling of sparse sinusoidal mixtures,'' in \emph{2018
  IEEE International Symposium on Information Theory (ISIT)}, 2018, pp.
  336--340.

\bibitem{bhandari2020unlimited}
------, ``On unlimited sampling and reconstruction,'' \emph{IEEE Transactions
  on Signal Processing}, 2020.

\bibitem{bk20}
A.~Bhandari and F.~Krahmer, ``Hdr imaging from quantization noise,'' in
  \emph{2020 IEEE International Conference on Image Processing (ICIP)}, 2020,
  pp. 101--105.

\bibitem{romanov2019above}
E.~Romanov and O.~Ordentlich, ``Above the {N}yquist rate, modulo folding does
  not hurt,'' \emph{IEEE Signal Processing Letters}, vol.~26, no.~8, pp.
  1167--1171, 2019.

\bibitem{bk19}
A.~Bhandari and F.~Krahmer, ``On identifiability in unlimited sampling,'' in
  \emph{2019 13th International conference on Sampling Theory and Applications
  (SampTA)}, 2019, pp. 1--4.

\bibitem{gbk19}
O.~Graf, A.~Bhandari, and F.~Krahmer, ``One-bit unlimited sampling,'' in
  \emph{ICASSP 2019 - 2019 IEEE International Conference on Acoustics, Speech
  and Signal Processing (ICASSP)}, 2019, pp. 5102--5106.

\bibitem{bkp21}
A.~Bhandari, F.~Krahmer, and T.~Poskitt, ``Unlimited sampling from theory to
  practice: Fourier-prony recovery and prototype adc,'' \emph{arXiv preprint
  arXiv:2105.05818}, 2021.

\bibitem{berger1971ratedistortion}
T.~Berger, \emph{Rate Distortion Theory: A Mathematical Basis for Data
  Compression}.\hskip 1em plus 0.5em minus 0.4em\relax Englewood Cliffs, NJ,
  USA: Prentice-Hall, 1971.

\bibitem{haykin2003least}
S.~S. Haykin, B.~Widrow, and B.~Widrow, \emph{Least-mean-square adaptive
  filters}.\hskip 1em plus 0.5em minus 0.4em\relax Wiley Online Library, 2003,
  vol.~31.

\bibitem{gray1998quantization}
R.~M. Gray and D.~L. Neuhoff, ``Quantization,'' \emph{IEEE Transactions on
  Information Theory}, vol.~44, no.~6, pp. 2325--2383, 1998.

\bibitem{lipshitz1992quantization}
S.~P. Lipshitz, R.~A. Wannamaker, and J.~Vanderkooy, ``Quantization and dither:
  A theoretical survey,'' \emph{Journal of the Audio Engineering Society},
  vol.~40, no.~5, pp. 355--375, 1992.

\bibitem{feuer1985convergence}
A.~Feuer and E.~Weinstein, ``Convergence analysis of {LMS} filters with
  uncorrelated {G}aussian data,'' \emph{IEEE Transactions on Acoustics, Speech,
  and Signal Processing}, vol.~33, no.~1, pp. 222--230, 1985.

\bibitem{aminikhanghahi2017survey}
S.~Aminikhanghahi and D.~J. Cook, ``A survey of methods for time series change
  point detection,'' \emph{Knowledge and information systems}, vol.~51, no.~2,
  pp. 339--367, 2017.

\bibitem{fawcett1997adaptive}
T.~Fawcett and F.~Provost, ``Adaptive fraud detection,'' \emph{Data mining and
  knowledge discovery}, vol.~1, no.~3, pp. 291--316, 1997.

\bibitem{leadbetter2012extremes}
M.~R. Leadbetter, G.~Lindgren, and H.~Rootz{\'e}n, \emph{Extremes and related
  properties of random sequences and processes}.\hskip 1em plus 0.5em minus
  0.4em\relax Springer Science \& Business Media, 2012.

\bibitem{benveniste2012adaptive}
A.~Benveniste, M.~M{\'e}tivier, and P.~Priouret, \emph{Adaptive algorithms and
  stochastic approximations}.\hskip 1em plus 0.5em minus 0.4em\relax Springer
  Science \& Business Media, 2012, vol.~22.

\bibitem{kamen1999wiener}
E.~Kamen and J.~Su, ``The {W}iener filter,'' in \emph{Introduction to Optimal
  Estimation}.\hskip 1em plus 0.5em minus 0.4em\relax Springer, 1999, pp.
  101--147.

\bibitem{banelli2000theoretical}
P.~Banelli and S.~Cacopardi, ``Theoretical analysis and performance of {OFDM}
  signals in nonlinear {AWGN} channels,'' \emph{IEEE Trans. on Communication},
  vol.~48, no.~3, pp. 430--441, 2000.

\end{thebibliography}

\end{document}